\documentclass[a4paper]{llncs}
\pdfoutput=1 

\usepackage{xspace}
\usepackage[english]{babel}
\usepackage[latin1]{inputenc}

\usepackage{gastex}
\usepackage[svgnames]{xcolor}
\usepackage{subfigure}
\usepackage{amsmath,amssymb}
\usepackage{enumerate}

\usepackage{url}

\usepackage{tikz}
\usetikzlibrary{shapes.misc}
\usetikzlibrary{calc}
\usetikzlibrary{automata}
\usetikzlibrary{backgrounds}
\usetikzlibrary{decorations.pathreplacing}
\usetikzlibrary{fadings}
\usetikzlibrary{automata,positioning,shapes.geometric,fit}

\tikzset{initial text={},
    every state/.style={circle,minimum size=.4cm,draw=blue!50,very thick,fill=blue!20},
    secret/.style={minimum size=.4cm,draw=red!50,very thick,fill=red!20,rectangle},
    node distance=1.5cm,on grid,auto,
    bend angle=65}

\def\malar{M\"alardalen University\xspace}
\def\ie{{i.e.},~}
\def\eg{{e.g.},~}
\def\st{{s.t.}~}

\def\unk{\bot} 
\def\start{i_1} 

\def\et{\mbox{\normalfont\textsf{time}}}
\def\stree{\mbox{\normalfont\textsf{tree}}}
\def\wcet{\normalfont\text{WCET}}

\def\iread{\textit{regR}}
\def\mread{\textit{memR}}
\def\iwrite{\textit{regW}}
\def\mwrite{\textit{memW}}
\def\flag{\textit{flags}}
\def\setNZ{\textit{SetStatusB}}
\def\cmpU{\textit{cmpU}}
\def\NDcmp{\textit{NDcmp}}
\def\update{\textit{update}}

\def\runs{\textit{Runs}}   

\newtheorem{fact}{Fact}  

\newcommand{\setB}{\mathbb B}

\def\calD{{\cal D}}
\def\calM{{\cal M}}

\def\calP{{\cal P}}
\def\calR{{\cal R}}

\def\calV{{\cal V}}

\def\endef{\ifmmode\squareforged\else{\unskip\nobreak\hfil
\penalty50\hskip1em\null\nobreak\hfil$\blacksquare$
\parfillskip=0pt\finalhyphendemerits=0\endgraf}\fi}

\def\true{\mbox{\textsc{true}}}
\def\false{\mbox{\textsc{false}}}

\usepackage{listings}

\definecolor{gris}{rgb}{0.3, 0.3, 0.3}

\lstdefinelanguage{AssemblerARM9}{%
basicstyle={\fontsize{6}{7}\ttfamily},
commentstyle={\color{gris}\it},
morekeywords={ldr,str,add,sub,b,bl,mov,cmp,movgt,ble,cmple,stmdb,ldmdb,ldm},
keywordstyle={\ttfamily  \textbf},%
sensitive=false,%
backgroundcolor=\color{blue!20!white},
rulecolor=\color{blue!50!white},
fillcolor=\color{blue!20!white},
texcl=true,
flexiblecolumns=true,
morecomment=[s][\it\color{gris}]{/*}{*/},
showstringspaces=true,
}

\renewcommand*\thelstnumber{${\the\value{lstnumber}}\!\!:$}
\newcommand{\lline}[1]{\the\value{#1}}

\makeatletter
\newcommand{\linelabel}[1]{%
\addtocounter{figure}{1}%
\immediate\write\@auxout{\string\newlabel{#1}{{\the\value{lstnumber}}{\thepage}%
{Line numbering}{figure.\thefigure}{}}}%
\addtocounter{figure}{-1}%
}
\makeatother

\newcommand{\FC}[1]{\textcolor{blue}{#1}}

\newcommand{\sem}[1]{[\![#1]\!]}

\title{Timed Games for Computing \\ Worst-Case Execution-Times}

\author{ Franck Cassez\thanks{Author supported by a Marie Curie
    International Outgoing Fellowship within the 7th
    European Community Framework Programme.}}

\institute{
  National ICT Australia \& CNRS \\ The University of New South Wales
  \\ Sydney, Australia }

\begin{document}

\maketitle
  
\thispagestyle{empty}

\begin{abstract} 
  In this paper we introduce a framework for computing upper bounds
  yet accurate WCET for hardware platforms with caches and pipelines.
  The methodology we propose consists of 3 steps: 1) given a program
  to analyse, compute an equivalent (WCET-wise) abstract program; 2)
  build a timed game by composing this abstract program with a network
  of timed automata modeling the architecture; and 3) compute the WCET
  as the optimal time to reach a winning state in this game.
  We demonstrate the applicability of our framework on standard
  benchmarks for an ARM9 processor with instruction and data caches,
  and compute the WCET with UPPAAL-TiGA. We also show that this
  framework can easily be extended to take into account dynamic
  changes in the speed of the processor during program execution.
\end{abstract}

\section{Introduction}
Embedded real-time systems are composed of a set of tasks (software)
that run on a given architecture (hardware).  These systems are
subject to strict timing constraints and these constraints must be
enforced by a scheduler.  Designing an effective scheduler is possible
only if some bounds are known about the execution times of each
task. For simple scheduling algorithms \eg non preemptive, the
knowledge of the \emph{worst-case execution-time} (WCET) is sufficient
to design a scheduler. For more complex scheduling algorithms with
preemption or shared resources, the WCET for each task might not give
rise to the WCET for the entire system though.  This is why most
critical embedded systems rely on a rather simple scheduling
algorithm.  Performance wise, determining tight bounds for WCET is
crucial as using rough over-estimates might either result in a set of
tasks being wrongly declared non schedulable or a lot of computation
time might be wasted in idling cycles and loss of energy/power.

\paragraph{\bfseries The WCET Problem.}
The execution time, $\et(p,d,H)$, of a program $p$, with input data
$d$ on the hardware $H$, is measured as the number of cycles of the
fastest component of the hardware \ie the processor.  Data take their
values in a finite domain $\calD$.  The program is given in binary
code or equivalently in the assembly language of the target
processor\footnote{When we refer to the ``source'' code, we assume the
  program $p$ was generated by a compiler, and refer to the high-level
  program (\eg in C) that was compiled into $p$.}.  The
\emph{worst-case execution-time} of program $p$ on hardware $H$ is
defined by:
 $$\wcet(p,H)=\sup_{d \in \calD} \et(p,d,H) \mathpunct.$$
The WCET problem asks the following: Given $p$ and $H$, compute
$\wcet(p,H)$.

\smallskip

In general, the WCET problem is undecidable because otherwise we could
solve the halting problem\footnote{Note this is true even for input
  data ranging over a finite domain, and can be proved using K\"onig's
  Lemma.}.  However, for programs that always terminate and have a
bounded number of paths, it is obviously (theoretically) computable.
Indeed the possible runs of the program can be represented by a finite
tree.
Notice that this does not mean that the problem is tractable though.

If the input data are known or the program execution time is
indepedent from the input data, the tree contains a single path and it
is usually feasible to compute the WCET.  Likewise, if we can
determine some input data that produces the WCET (this might be as
difficult as computing the WCET), we can compute the WCET on a
single-path program.

\smallskip

If is not often the case that the input data are known or that we can
determine an input that produces the WCET.  Rather the (values of the)
input data are unknown, and the number of paths to be explored might
be extremely large: for instance, for a Bubble Sort program with $100$
data to be sorted, the tree representing all the runs of the
(assembly) program on all the possible input data has more than
$2^{50}$ nodes.  Although symbolic methods (\eg using BDDs) can be
applied to analyse some programs with a huge number of states, they
will fail to compute the exact WCET on Bubble Sort by exploring all
the possible paths.

\smallskip

Another difficulty of the WCET problem stems from the more and more
complex architectures embedded real-time systems are running on.  They
usually feature a multi-stage \emph{pipeline} and a fast memory
component like a \emph{cache}, and they both influence in a
complicated manner the WCET.  It is then a challenging problem to
determine a precise WCET even for relativey small programs running on
complex architectures.

\paragraph{\bfseries Methods and Tools for the WCET Problem.}
The reader is referred to~\cite{wcet-survey-2008} for an exhaustive
presentation of the WCET computation techniques and tools.  There are
two main classes of methods for computing WCET.
\begin{itemize}
\item Testing-based methods. These methods are based on experiments
  \ie running the program on some data, using a simulator of the
  hardware or the real platform. The execution time of an experiment
  is measured and, on a large set of experiments, a maximal and
  minimal bound can be obtained. The maximal bound computed this way
  is \emph{unsafe} as not all the possible paths have been explored.
  These methods might not be suitable for safety critical embedded
  systems but they are versatile and rather easy to implement.

  RapiTime~\cite{rapitime} (based on pWCET~\cite{pWCET}) and
  Mtime~\cite{mtime} are measurement tools that implement this
  technique.

\item Verification-based methods.  These methods often rely on the
  computation of an \emph{abstract} graph, the control flow graph
  (CFG), and an abstract model of the hardware.  Together with a
  static analysis tool they can be combined to compute WCET.  The CFG
  should produce a super set of the set of all feasible paths. Thus
  the largest execution time on the abstract program is an upper bound
  of the WCET.  Such methods produce \emph{safe} WCET, but are
  difficult to implement. Moreover, the abstract program can be
  extremely large and beyond the scope of any analysis. In this case,
  a solution is to take an even more abstract program which results in
  drifting further away from the exact WCET.

  Although difficult to implement, there are quite a lot of tools
  implementing this scheme: Bound-T~\cite{bound-T},
  OTAWA~\cite{otawa}, TuBound~\cite{tubound},
  Chronos~\cite{chronos}, 
  SWEET~\cite{sweet-2003} and
  aiT~\cite{aiT,wcet-ai-aswsd-ferdinand-04} are static analysis-based
  tools for computing WCET.
\end{itemize}

The verification-based tools mentioned above rely on the construction
of a control flow graph, and the determination of loop bounds.  This
can be achieved using user annotations (in the source code) or
sometimes infered automatically.  The CFG is also annotated with some
timing information about the cache misses/hits and pipeline stalls,
and paths analysis is carried out on this model \eg by Integer Linear
Programming (ILP).  The algorithms implemented in the tools use both
the program and the hardware specification to compute the CFG fed to
the ILP solver.  The architecture of the tools themselves is thus
monolithic: it is not easy to adapt an algorithm for a new processor.
This is witnessed by \emph{WCET'08 Challenge
  Report}~\cite{wcet-chal-report-08} that highlights the difficulties
encountered by the participants to adapt their tools for the new
hardware in a reasonable amount of time.

\paragraph{\bfseries WCET and Model-Checking.}

Surprisingly enough, only a few tools use model-checking techniques to
compute WCET.  Considering that ($i$) modern architectures are
composed of \emph{concurrent} components (the stages of the pipeline,
caches) and ($ii$) these components \emph{synchronize} and
synchronization depends on \emph{timing constraints} (time to execute
in one stage of the pipeline, time to fetch a data from the cache),
formal models like \emph{timed automata}~\cite{AD94} and
state-of-the-art \emph{real-time model-checkers} like
UPPAAL\cite{uppaal-sttt-97,uppaal-40-qest-behrmann-06} appear
well-suited to address the WCET problem.

It has previously been claimed~\cite{Wilhelm-04} that
\emph{model-checking} was not adequate to compute WCET, but this
statement has since been revised.  In~\cite{wcet-cav-metzner-04},
A.~Metzner showed that model-checkers could well be used to compute
safe WCET on the CFG for programs running on pipelined processors with
an instruction cache.

In~\cite{hubert-wcet-09}, B.~Hubert and M.~Schoeberl consider Java
programs and compare ILP-based techniques with model-checking
techniques using the model-checker UPPAAL.  Model-checking techniques
seem slower but easily amenable to chan\-ges (in the hardware model).
The recommendation is to use ILP tools for large programs and
model-checking tools for code fragments.

More recently, the TASM toolset~\cite{tasm-cav-07} (M. Ouimet \&
K. Lundqvist) has been used to compute WCET with UPPAAL: the TASM
machine is a high level machine not featuring pipelining nor caches
and computing the WCET amounts to finding the longest path (timewise)
in a timed automaton that specifies a tasks.

Another use of timed automata (TA) and the model-checker UPPAAL for
computing WCET on pipelined processors with caches is reported
in~\cite{metamoc-2009}.  The framework METAMOC described
in~\cite{metamoc-2009} (A.~E.~Dalsgard \emph{et al.}) consists in: 1)
computing a flow graph (FG) from a binary program, 2) composing this
FG with a (network of timed automata) model of the processor and the
caches.  Computing the WCET is then reduced to a safety (or dually a
reachability) property $\mathsf{AG} \; (\text{Time} \leq k)$ (reads
``on all paths, the variable \text{Time}, global time, is less than
$k$'') that can be checked with UPPAAL.

\medskip

The previous framework is extremelly elegant yet has some
shortcomings.  Out of the $15$ programs\footnote{The benchmarks
  contain 35 programs. In~\cite{metamoc-2009}, only 14 programs can be
  analysed with a concrete instruction cache and 7 with a concrete
  instruction and data cache.} of the \malar benchmarks only $7$ can
be analysed with a concrete instruction and data cache (Table~.6.1,
page~84 in~\cite{metamoc-2009}).  It is also surprising that some
single-path programs could not be analysed with concrete caches. The
tool chain relies on a value analysis tool which fails on $3$ of the
$15$ programs.  It requires a specialised version of UPPAAL (not
available) to avoid a binary search for computing the WCET.

\paragraph{\bfseries Our Contribution.}
In this paper we use \emph{timed game automata} (TGA) and
UPPAAL-TiGA~\cite{tiga-cav-07} (UPPAAL for timed games) to compute
WCET.  We model the WCET problem as a two-player timed game.
Intuitively Player~1 is the program, and Player~2 is in charge of
deciding the outcome of the \emph{comparison} instructions (\eg
\texttt{cmp, tst} which set the branching conditions) that depend on
the input data.  As the choice of the input data is not controllable
by Player~1, we obtain a two-player game.  The problem we solve on
this game is an \emph{optimal time reachability problem}:
\begin{center}
  ``What is the optimal time for Player~1  to reach the end
  of the program~?''
\end{center}
What is similar to the previously mentioned
approach~\cite{metamoc-2009} (A.~E.~Dalsgard \emph{et al.}) is the
timed automata models for the caches\footnote{Note that a similar
  model is reportedly due to A.~P.~Ravn in~\cite{hubert-wcet-09}.} and
pipeline stages \ie the model of the architecture, but we use a
totally different model for the program.  We propose a new and very
compact encoding of the program and pipeline stages' states which
enables us to compute the WCET for $13$ out of the previous $15$
programs\footnote{Say why $2$ fails ...} (see Table~\ref{tab-results},
page~\pageref{tab-results}).  Moreover, compared to METAMOC that uses
a computer with 32GB RAM, we can compute the results on a laptop
computer (2Ghz Dual Core, 2GB RAM) within a few seconds.  Using timed
games instead of timed automata is also a major difference: the
on-the-fly algorithm~\cite{cassez-concur-05} implemented in
UPPAAL-TiGA is different from the one running in UPPAAL, and it can
also compute the \emph{optimal time} (in the presence of adversary) to
reach a designated state. Thus we do not need to do a binary search or
use a tailored version of UPPAAL to compute the results.

\smallskip

We also show that taking into account processor speed variations is
easy in our framework. This can be important as it is possible to
adjust the speed of the processor depending on the program to be run.
For some programs, the saved power can be upto 22\% (see
Table~\ref{tab-results}).


\smallskip

\noindent The advantages of our approach are many-fold
(METAMOC~\cite{metamoc-2009} shares 1--3):
\begin{enumerate}
\item it is very easy to implement as it consists of two separate and
  independant phases: 1) computation of a model of the program to be
  analysed; this only requires a (formal) semantics of the assembly
  language of the target processor\footnote{In contrast, the
    verification-based tools would need a description of the hardware
    to compute the CFG.}; 2) computation of the WCET with UPPAAL-TiGA
  and the models for the caches, pipelines which specify the timing
  features. A model of a cache (\eg always miss or FIFO) can be
  substitued by changing the cache component only (no need to
  recompute the model obtained in phase 1).
\item the design of the models for pipeline stages and caches can be
  stressed by simulating some simple samples programs; this enables us
  to get more confidence in the model of the hardware as this is not
  hidden in the analysis algorithm; this is especially important for
  concurrent architectures like pipelined processors that can be hard
  to describe;
\item UPPAAL or UPPAAL-TiGA can be used to simulate the program on the
  architecture. It is thus a quick way of obtaining a simulator for a
  given hardware;
\item we do not require annotations. Instead, we run a simulation of
  the program with some given bounds on the number of branching or a
  maximal number of states. If too many branchings are encountered,
  the user is required to provide a constraint for the corresponding
  instruction in the program to remove some infeasible paths;
\item we solve an \emph{optimal time reachability problem} on the
  program $p$ of the form: ``what is the optimal time to enforce
  \emph{termination} of program $p$ ?''.  This at once 1) proves that
  $p$ terminates on every input data, and 2) computes the WCET.
  This could not be achieved in METAMOC~\cite{metamoc-2009} as the
  UPPAAL model contains priorities and deadlock freedom cannot be
  checked on models with priorities: thus if the safety property
  $\mathsf{AG} \; (\text{Time} \leq k)$ is satisfied, it does not mean
  that no deadlocks occurred; the deadlocks could be due to a flaw in
  the design of the pipeline model but in any case, it does not give a
  safe bound for the WCET as deadlocks have not been excluded.
\item it is easy to add \emph{power} related constraints in the model
  \eg processor speed variations;
\item we also show that not every program instruction is worth
  simulating and some \emph{abstraction} on the effect of some
  instructions can be safely done. 
  For example, in the Fibonnacci program,
  the content of the variable with the result is irrelevant for the
  computation of the WCET. It does not influence any branching nodes.
  We show how to check that an abstract program is \emph{equivalent}
  to a concrete one and examplify this on some of the benchmarks from
  \malar.
\end{enumerate}

\paragraph{\bfseries Outline of the Paper.}
In Section~\ref{sec-concrete-abs}, we briefly introduce the ARM9
architecture and the assumptions we make on the assembly programs to
be analysed.  Section~\ref{sec-games} describes how to encode an
assembly program with non-deterministic choices into a game.  In
Section~\ref{sec-architecture} we give the timed automata models of
the architecture we use to compute the
WCET. Section~\ref{sec-tool-chain} gives an overview of the tool chain
we propose and the components (compiler) we have designed together
with some comments on the case studies presented in
Table~\ref{tab-results}.


\section{Concrete and Abstract Programs}
\label{sec-concrete-abs}
\paragraph{\bfseries Program, Registers, Memory.}
A \emph{program} $p$ is a list of instructions $p=i_1,i_2,\cdots,i_k$
and $i_1$ is the initial instruction.  The control usually goes from
intruction $i_k$ to $i_{k+1}$ except for branching intructions that
give the next instruction $i_j$ to be performed.  Each instruction
performs some basic operations (arithmetic, logic, memory load or
store, branching) and has a duration which gives the amount of time it
takes in each stage of the pipeline of the processor\footnote{A
  particular case is a processor with one stage.}.  We assume the
duration is independant from the content of the operands of the
instructions\footnote{This is not always the case as for instance the
  duration of the instruction \texttt{mull} (multiplication on long
  integers) on the AMRM9 depends on how large one of the operand
  is. However, we can always take the longest duration to obtain a
  safe upper bound of the WCET.}.  In the sequel we use the variable
$\iota$ to denote an instruction of $p$.

 \smallskip

 The hardware on which $p$ runs has a pool of \emph{registers}
 (different from the main memory and the caches).  We let
 $\calR=\{r_0,\cdots,r_k\}$ be the set of registers.  For example on
 the ARM9~\cite{arm9-ref} processor there are $16$ registers.  A
 designated register \textbf{pc} contains the program counter and
 points to the next instruction to be performed (register $15$ on the
 ARM9).

\smallskip

We let $\calM=\{m_1,m_2,\cdots,m_n\}$ be the set of memory cells'
addresses used by the program (we assume the program can access
$\calM$).
The content of the memory cells and registers is in a finite domain
$\calD$ (\eg $32$ bit integers).

\paragraph{\bfseries Semantics.}
When program $p$ runs on input data $d$, it generates a computation
that changes the values of the registers and memory cells.

A state (of the computation of $p$) is given by a mapping $v: \calR
\cup \calM \rightarrow \calD$ and we let $\calV$ be the set of states.

Performing an instruction results in a state change, and is
deterministic. Given an instruction $\iota$ ($\iota$ including the
operands and can be thought of as the \emph{code} of an assembly
instruction), the semantics of $\iota$ is a mapping $\sem{\iota}:
\calV \xrightarrow{} \calV$.

As the program counter is in one of the registers, the semantics of a
program $p$ is completely determined by the current state of the
computation.  From a state $v$, the next state (in the computation of
$p$) is $v'$ and we denote this $v \xrightarrow{} v'$. $v'$ is given
by $\sem{\iota}(v)$ where and $\iota=i_\mathbf{pc}$ (we use
\textbf{pc} both for the register and the content of this register to
avoid hefty notations).

For branching instructions, the control is determined by the
\emph{status bits} and we assume there are also part of the
\textbf{pc} register.
\begin{remark}
  We assume \textbf{pc} is incremented by $1$ after each instruction
  (except for branching instruction). In an actual computer, it is
  incremented by the \emph{word size} but these details are irrelevant
  at this stage.
\end{remark}



\paragraph{\bfseries Side Effects of an Instruction.}

Each instruction reads from and writes to some subset of registers.
We let $\iread(\iota)$ (resp. $\iwrite(\iota)$) be the set of ``read
from'' (resp. ``written to'') registers for instruction $\iota$.

Each instruction can also read or write to main memory cells.  We let
$\mread(\iota)$ (resp. $\mwrite(\iota)$) be the set of memory cells
addresses read from (resp. written to) by instruction $\iota$.


\noindent\begin{minipage}[c]{.48\textwidth}
\thicklines
\vspace*{.18cm}
\begin{lstlisting}[language=AssemblerARM9,numbers=none,caption={Binary Search Program},label={lis-bs}]
00000000 <main>:
 0:   e3a00009  mov     r0, #9  ; 0x9
 4:   eaffffff  b       8 <binary_search>

00000008 <binary_search>:
 8:   e92d4030  stmdb   sp!, {r4, r5, lr}
 c:   e59f4040  ldr     r4, [pc, #64] ;
10:   e3a0e000  mov     lr, #0  ; 0x0
14:   e3a0c00e  mov     ip, #14 ; 0xe
18:   e3e05000  mvn     r5, #0  ; 0x0
1c:   e08e300c  add     r3, lr, ip
20:   e1a020c3  mov     r2, r3, asr #1
24:   e7943182  ldr     r3, [r4, r2, lsl #3]
28:   e0841182  add     r1, r4, r2, lsl #3
2c:   e1530000  cmp     r3, r0            %\textbf{/ eq le /}%
30:   05915004  ldreq   r5, [r1, #4]
34:   024ec001  subeq   ip, lr, #1      ; 0x1
38:   0a000001  beq     44 <binary_search+0x3c>
3c:   c242c001  subgt   ip, r2, #1      ; 0x1
40:   d282e001  addle   lr, r2, #1      ; 0x1
44:   e15e000c  cmp     lr, ip              %\textbf{/ le /}%
48:   c1a00005  movgt   r0, r5
4c:   dafffff2  ble     1c <binary_search+0x14>
50:   e8bd8030  ldmia   sp!, {r4, r5, pc}
54:   00000158  andeq   r0, r0, r8, asr r1
\end{lstlisting}
\end{minipage}\hfill
\begin{minipage}[c]{0.49\linewidth}
  An example of an assembly program is given in Listing~\ref{lis-bs}.
  This program performs a binary search on an array of 14 elements.
  Line~24 loads register \textsf{r3} with a value of the array at
  address $v(\mathsf{r4})+(v(\mathsf{r2})*8)$. As we do not know the
  values of the array, the value of \textsf{r3} is unknown after this
  instruction. \textsf{r0} contains the value we are looking for, and
  is also
  unknown\refstepcounter{footnote}\footnotemark[\value{footnote}].  As
  a consequence, the comparison of line~2c is undetermined as the
  value of \textsf{r3} in unknown. The outcome of the comparison is
  used later in conditional instructions (\eg \texttt{ldreq r5, [r1,
    \#4]} and \texttt{subgt ip,r2,\#1}) and branching instructions
  \texttt{beq 44}. \hfill Two \emph{status bits} are needed to 
\end{minipage}
\noindent 
encode the  result of the comparison at line~24: whether \textsf{r3} is ``lower
  or equal'' than \textsf{r0} and whether \textsf{r3} is ``equal'' to
  \textsf{r0}. This is indicated by the two
predicates \textsf{eq} and \textsf{le} between \textrm{/ \ldots /}.
\footnotetext{In the actual program it is $9$ but it does not change
  the execution tree of the program.}   
The address of the memory cell referenced at line~24 is determined by
the previous outcomes of the comparison instruction at line~2c.


\paragraph{\bfseries Runs.}

A \emph{run} of program $p$ from state $v_0$ (initial value of the
input data) is the (unique) sequence of instructions performed by $p$
from $v_0$:
\[
\rho(p,v_0)= \iota_1 \quad \cdots \quad \iota_k \quad \cdots \quad \iota_n
\]
with $\iota_1=i_1$.  The length of the run $\rho(p,v_0)$ is
$|\rho(p,v_0)|=n$.  We assume that every run terminates, and that
moreover, given $p$, there exists a contant $K_p$ \st $\forall v \in
\calV, |\rho(p,v_0)| \leq K_p$.  Intuitively, this means that all
loops are bounded, and it implies that there is no run which
encounters twice the same state.

The state after the subsequence $\iota_1 \ \cdots \ \iota_k$ is
determined by the composition of the semantics function of each
instruction.  If $v_{j}$ is the state after instruction $\iota_j$ then
$v_{j+1}=\sem{\iota_{j+1}}(v_j)$, and $v_0$ is the initial state.

\paragraph{\bfseries Execution Time of a Run.}
If each instruction was performed one after the other, the
execution-time of a run would be the sum of the execution times of
each instruction.

\noindent On pipelined architectures with caches, the execution-time
solely depends on:
\begin{enumerate}
\item the subsequences of instructions: pipeline \emph{stalls} can
  occur, for instance because one instruction (\eg in the execute
  stage) reads a register written to by the instruction in the next
  stage (\eg memory stage).
\item the time to read or write a memory cell: instructions that
  require memory transfers (load and store) might take different
  durations if a \emph{cache} is used, depending on whether the memory
  cell is already in the cache of not.
\end{enumerate}
We let $H$ denote the architecture of the system. $H$ refers to the
pipeline structure and timing specifications, the cache initial state,
size, replacement policy and timing specifications, and the timing
specifications of the main memory.
The \emph{execution-time} of a run $\rho$ is completely determined by:
\begin{itemize}
\item the architecture $H$, 
\item the duration of each instruction of $\rho$ in each stage of the
  pipeline,
\item the registers read from and written to, and memory cells read
  from or written to by each instruction of $\rho$.
\end{itemize}
The \emph{duration} of a run $\rho$ on architecture $H$ is denoted
$\et_H(\rho)$.  This function might be rather complex but is yet
well-defined.

To formalize the previous informal definition, assume the architecture
$H$ is fixed.  Let $\rho=\iota_1 \, \cdots \, \iota_n$ and
$\rho'=\iota'_1 \, \cdots \, \iota'_n$ be two runs of program $p$.  We
say that $\rho$ and $\rho'$ are (time-wise) \emph{$H$-equivalent} and
write $\rho \approx_H \rho'$ if for each $1 \leq k \leq n$:
\begin{itemize}
\item the duration of $\iota_k$
  in each stage of the pipeline is the same as the duration
  of $\iota'_k$;
\item the registers used as operands and memory cells referenced
  are also the same: $\phi(\iota_k)=\phi(\iota'_k)$ for
  $\phi \in \{ \iread,\iwrite,\mread,\mwrite\}$.
\end{itemize}
\begin{fact}\label{fact-equiv}
  If $\rho \approx_H \rho'$ then $\et_H(\rho)=\et_H(\rho')$.
\end{fact}
The \emph{worst-case execution-time} for program $p$ on architecture
$H$ is given by:
\[
\wcet(p,H) = \max_{v_0 \in \calD} \et_H(\rho(p,v_0))\mathpunct.
\]

\paragraph{\bfseries Timing Anomalies.}

\emph{Timing anomalies}~\cite{wcet-survey-2008} can occur because of
the complex architecture of the hardware $H$.  The term refers to
counter-intuitive observations in the sense that larger \emph{local}
execution-times may not result in larger \emph{global}
execution-times.  \emph{Pre-fetching} instructions can lead to such
observations on some processors. This can also be observed on complex
pipeline architectures (\eg \emph{out-of-order} execution of
instructions).

\bigskip

On architectures that do not exhibit \emph{timing anomalies}, the
function $\et_H$ is in some sense \emph{monotonic}.

For instance an achitecture $H_\mu$ with an ``always miss'' cache (or
equivalently no cache) will produce a WCET which is always greater
than on an architecture $H$ with a cache of size more than $1$.  As we
consider worst-case execution-time, a \emph{random} replacement policy
for a cache is equivalent to an ``always miss'' cache.
Let $H_{r}$ denote a cache with random replacement policy, and $H$ a
regular cache (LRU, FIFO, semi-random replacement policy).  The
following holds:
\begin{fact}
  $\wcet(p,H) \leq \wcet(p,H_\mu) = \wcet(p,H_r).$
\end{fact}
This implies that an over-approximation of $\wcet(p,H)$ can always be
obtained using an equivalent architecture $H'$ with an ``always miss''
cache.

\smallskip

The same remark applies for the pipeline of architecture $H$.  If $H'$
is the same as $H$ with larger durations for each instruction at each
stage, then $\wcet(p,H)\!\!\leq \wcet(p,H')$. If a pipeline \emph{stall}
in $H$ implies a pipeline stall in $H'$ for every program and every
input data, then $\wcet(p,H) \leq \wcet(p,H')$.

Another interesting case is when a \emph{branch} instruction is
executed.  If it is not a loop, the program fragment has a diamond
shape: both branches join at some future point in the computation.  If
the local worst-case execution time is obtained by taking one side of
the branch instruction, we can safely ignore the other side as it does
not contribute (more) to the global worst-case execution-time.

The framework of this paper does handle timing anomalies, but some
abstractions defined below are not safe for architecture exhibiting
timing anomalies.

\paragraph{\bfseries Abstractions.}
In this section we introduce some simple abstractions that can be made
on a program $p$.  The aim of this abstraction is to reduce the space
needed to encode the state of the computation.  We examplify the
usefulness of these abstractions on some benchmarks programs from
\malar.

\begin{figure}[htbp]
\centering
\begin{minipage}{.44\textwidth}\centering \thicklines
\begin{lstlisting}[caption={C Program},label={fib-C}]
int fib(int n)
{
  int i,Fnew,Fold,temp,ans;
    Fnew=1;Fold = 0;
    for(i=2;i<=30 && i<=n; i++)
    {
      temp=Fnew;
      Fnew=Fnew + Fold;
      Fold=temp;
    }
    ans=Fnew;
  return ans;
}
\end{lstlisting}
\end{minipage}
\hspace*{.3cm}
\begin{minipage}{.44\textwidth}\centering \thicklines
\begin{lstlisting}[language=AssemblerARM9,numbers=none,caption={Assembly Code},label={fib-as}]
   0: mov    r2, #2       ; 0x2
   4: cmp    r2, r0
   8: mov    ip, r0
   c: mov    r0, #1       ; 0x1
  10: mov    r1, #0       ; 0x0
  14: movgt  pc, lr
  18: add    r2, r2, #1   ; 0x1
  1c: mov    r3, r0
  20: cmp    r2, #30      ; 0x1e
  24: cmple  r2, ip
  28: add    r0, r0, r1
  2c: mov    r1, r3
  30: ble    18 <fib+0x18>
  34: mov    pc, lr
\end{lstlisting}
\end{minipage}
\caption{Fibonacci Program.}
 \label{fig-fibo}
\end{figure}
Listing~\ref{fib-C} (Fig.\ref{fig-fibo}) gives a C function computing
the Fibonacci number $n$.  Its assembly language version is given in
listing~\ref{fib-as}.  The control flow of the assembly version is
controlled by lines~20, 24 and~30: register \textsf{r2} contains the
loop variable $i$ and is incremented at each round.  Lines~c, 10, 1c,
28 and 2c are not contributing to the program control flow.  If we are
only interested in the \emph{execution-time} of this program, their
\emph{effects} can be safely abstracted away.  We can replace them by
equivalent instructions that modify only the \textbf{pc} register,
with the same read/written registers (and memory cells if it happens
to be a load/store instruction).  For instance, instruction
\textsf{mov} at line c, can be replaced by an \emph{abstract}
instruction \textsf{mov$^a$} with:
\begin{itemize}
\item $\sem{\textsf{mov$^a$}}(v)=v'$ with $v'(r)=v(r)$ for each
  register different from \textbf{pc} 
  and $v'(\mathbf{pc})=v(\mathbf{pc})+1$;
\item the duration of \textsf{mov$^a$} in each stage of the
  pipeline is the same as \textsf{mov};
\item the registers read from/written to by \textsf{mov$^a$} at line c
  are the same as the ones read from/written to by instruction
  \textsf{mov} at line c.
\end{itemize}

In the end, we can abstract away the values of registers \textsf{r0},
\textsf{r1} and \textsf{r3} and assume they are always $0$ as no
abstract instruction will modify them.  The WCET of the abstracted
program will be exactly the same as the concrete one.

The goal of this abstraction is to reduce the space needed to encode a
state of the computation. Instead of encoding $7$ registers, only $4$
are relevant for the computation of the WCET.

\smallskip

A valid abstract program must simulate the execution tree of the
concrete program.  To be equivalent WCET-wise to the concrete program,
it should also preserve the addresses of the referenced memory cells
to ensure that cache hits/misses are preserved.

\bigskip

To formalize the previous notions, we first define \emph{critical}
instructions.
A \emph{critical} instruction is an instruction that:
\begin{itemize}
\item[($i$)] either sets some status bits; it can be a comparison or
  test (\eg \textsf{cmp, tst}) or an arithmetic instruction with the
  ``s'' flag on the ARM9 (\eg a subtraction
  \textsf{subs~r2,~r2,~\#1});
\item[($ii$)] or an instruction that references a memory cell \eg
  \textsf{ldr r0,~[r2,~r3 lsl \#2]} (load register \textsf{r0} with
  the content of memory cell $\mathsf{r2} + (\mathsf{r3} \times 4)$).
\end{itemize}
Next we define \emph{abstract} instructions.  As examplified for the
\textsf{mov} instruction at line~c previously, given an instruction
$\iota$, the \emph{abstracted instruction} $\iota^a$ is defined
by:
\begin{itemize}
\item the semantics of $\iota^a$ is $\sem{\iota^a}(v)=v'$
  with $v'(x)=v(x)$ for each register $x$ different from \textbf{pc} and
  each memory cell $x$ in $\calM$, and $v'(\mathbf{pc})=v(\mathbf{pc})+1$;
\item the duration of $\iota^a$ in each stage of the pipeline is
  the same as the duration of $\iota$;
\item the registers read from/written to by $\iota^a$ are the
  same as the ones read from/written to by instruction $\iota$:
  $\phi(\iota)=\phi(\iota^a)$ for
  $\phi \in \{\iread,\iwrite,\mread,$ $\mwrite\}$.
\end{itemize}
Let $p^a=i_1^a \cdots i_n^a$ be the abstract program that corresponds
to $p=i_1 \cdots i_n$.  An \emph{abstraction mapping} $\alpha$ is a
mapping that associates with each (concrete) instruction $\iota$ of
$p$, either $\iota$ (identity) or $\iota^a$ ($\alpha$ determines
whether $\iota$ is abstracted or not).  We write $\iota^\alpha$ for
$\alpha(\iota)$.

Let $\rho(p,v_0)=\iota_1\iota_2 \cdots \iota_k$ be a run of $p$ from
$v_0$ and $\rho(p^\alpha,v_0)=\iota^\alpha_1\iota^\alpha_2 \cdots
\iota^\alpha_k$ the corresponding $\alpha$-abstracted run.  Let
$I_c(p,v_0)\subseteq \{1,2,\cdots,k\}$ be the set of indices \st $j+1
\in I_c(p,v_0) \iff \iota_{j+1}$ is a critical instruction in
$\rho(p,v_0)$. Let $v_{j}$ be the state after executing instruction
$j$ in $\rho(p,v_0)$ and $v_{j}^\alpha$ be the state after executing
abstract instruction $j$ in $\rho(p^\alpha,v_0)$.

The following Lemma states that, if the values of the registers read
from/written to by any critical instruction (in $\rho(p,v_0)$), are
equal to the values of the same registers in the abstract execution,
the execution time of the concrete and abstract run is the same.

\begin{lemma}\label{lem-1}
  If $\forall j+1 \in I_c(\rho(p,v_0))$, $v_j(r)=v^\alpha_j(r)$ for
  each $r \in \iread(\iota_{j+1}) \cup \iwrite(\iota_{j+1})$ then
  $\et_H(\rho(p,v_0))=\et_H(\rho(p^\alpha,v_0))$.
\end{lemma}
\begin{proof}
  If the values of the operand registers of each critical instruction
  $\iota_j$ are the same in the concrete and abstract runs before
  performing $\iota_j$ and $\iota_j^\alpha$,  then:
  \begin{enumerate}
  \item the status bits that are set by the critical instruction
    have the same values in the concrete and abstract state;
  \item the addresses of the memory cells referenced by the
    instruction are the same in the concrete and abstract run. 
  \end{enumerate}
  The concrete and abstract run are thus $H$-rquivalent, \ie
  $\rho(p,v_0) \approx_H \rho(p^\alpha,v_0)$.  By
  Fact~\ref{fact-equiv}, it follows that
  $\et_H(\rho(p,v_0))=\et_H(\rho(p^\alpha,v_0))$. \qed
\end{proof}
If Lemma~\ref{lem-1} holds for each run $\rho(p,v_0)$ with $v_0 \in
\calD$, we say that $p$ and $p^\alpha$ are $H$-equivalent and write $p
\approx_H p^\alpha$.  In this case, by definition of the WCET, we
have:
\begin{lemma} \label{lem-2}
  If $p \approx_H p^\alpha$ then $\wcet(p,H) = \wcet(p^\alpha,H)$.
\end{lemma}
 
\paragraph{\bfseries Context Independence.}
As we cannot simulate $p$ for every input data, we assume that the
initial values of these data can be arbitrarily chosen.  To formalize
this, we use an extended domain for the values of the registers and
memory cells: $\calD \cup \{\unk\}$ where $\unk$ is a special
\emph{unknown} value.  At the beginning of the computation, every
register (except \textbf{pc}) and memory cell has its value set to
$\unk$.  The initial state is thus $v_0$ with $v_0(x)=\unk$ for $x \in
(\calR \setminus \{\mathbf{pc}\}) \cup \calM$ and
$v(\mathbf{pc})=\start$ where $\start$ is the address of the first
instruction of program $p$.

We assume that for each program $p$, the addresses of the memory cells
referenced during the course of the execution of the program, only
depend on the current state and are independent from the input data
values. By this, we mean that the address referenced at each point in
a run of a program is determined by some registers values that are
known. These values may depend on the actual content of some memory
cells because they influence the branching instructions, but once a
branch is chosen, the addresses can be computed. An example is a
binary search program: we have to determine wether a sorted array $v$
contains a value $s$. The search continues as long as $s$ has not been
found.

The semantics of each instruction (next state) is extended to the
extended domain $\calD \cup \{\unk\}$ as follows:
\begin{itemize}
\item for arithmetic and logical instructions, the value of the result
  of an instruction is $\unk$ if the value of one of the operands is
  $\unk$;
\item for instructions that set the status bits, there might be more
  than one next state; if one operand is $\bot$, the next states are
  given by all the possible values of the status bits;
\item for memory transfer instructions (load, store with addresses in
  $\calM$) the result in memory or register is always $\bot$.
  Nevertheless, for transfers involving the \emph{stack} (a subset of
  the addresses in $\calM$), we keep track of the values pushed or
  popped. The stack is quite often used on call/return of a function,
  and abstracting the content of the stack would result in some
  infeasible paths, or even to references to forbidden memory cells.
\item for branching instructions, there is one next state determined
  by the value of the target (unconditional branching) or by the
  status bits (conditional branching).
\end{itemize}
From the previous extended definitions, there might be more than one
run from the initial extended state $v_0$.  We denote $p_\unk$ the
non-deterministic program that corresponds to $p$ on the extended
domain. The semantics of $p_\unk$ is a \emph{tree}, $\stree(p_\unk)$
where the branches correspond to the choices of the status bits when
required. Note that this tree might be unbounded.

An important property of this tree, is that if $\rho(p,v_0)$ is a run
of $p$ on input data $v_0$, there is a path $\rho'$ in
$\stree(p_\unk)$ that satisfies $\rho(p,v_0) \approx_H \rho'$.
Moreover, as we assume that the number of steps when running $p$ is
bounded by $K_p$, we can safely truncate the tree $\stree(p_\unk)$ and
prune all nodes that are more than $K_p$ steps apart from the root.
Let $\runs(p_\unk)$ denote the set of rooted paths in the tree
$\stree(p_\unk)$.  We assume $\stree(p_\bot)$ has depth at most $K_p$.
Let
$$\wcet(p_\bot,H)=\max_{\rho \in \runs(p_\unk)} \et_H(\rho)\mathpunct .$$
As every run of $p$ is simulated by a run $p_\unk$, we have:
\[
\wcet(p,H) \leq \wcet(p_\bot,H) \mathpunct .
\]

Moreover, we can also define an abstract version, $p_\bot^\alpha$, of
$p_\bot$, given an abstraction mapping $\alpha$. The definitions are
extended to te extended domain.
As before we have:
\begin{lemma}\label{lem-3}
  If $p_\bot \approx_H p_\bot^\alpha$, then $\wcet(p_\bot,H) =
  \wcet(p_\bot^\alpha,H)$.
\end{lemma}

Combining Lemma~\ref{lem-2} and Lemma~\ref{lem-3}, we have:
\begin{lemma}
  If $p_\bot \approx_H p_\bot^\alpha$, $\wcet(p,H) \leq
  \wcet(p_\bot^\alpha,H)$.
\end{lemma}

\paragraph{\bfseries Checking that $p^\alpha \equiv_H p$.}
Checking whether $p_\bot \approx_H p_\bot^\alpha$ can be done by
building a \emph{synchronized product} of $p_\bot$ and $p_\bot^\alpha$
and checking wether each state preceeding a critical instruction satisfies
the condition of Lemma~\ref{lem-1}.

This is implemented in our framework (see
Fig.~\ref{fig-tool-chain}) by generating a C++ file that
performs this check.

Table~\ref{tab-results}, column \emph{Abs} gives the ration of
abstracted instructions for some programs (when we have chosen to
abstract away some instructions).  For some programs (\texttt{matmult}
and \texttt{jfdcint}) the number of abstracted instructions is rather
high. This indicates that the control flow is quite simple and
governed by a small number of instructions.

Notice that this abstraction does not change the WCET of the program.

\section{From Programs to Games}
\label{sec-games}
In this section we describe how to encode an assembly program into a
 game.  The encoding can be applied to any assembly language but
we give examples for the ARM9 processor.

Given a program $p$, we define a two-player game to model the runs of
$p_\bot$ defined in the previous section.  Player~1 executes the
instructions of $p_\bot$. The role of Player~2 is to set the values of
the status bits when an instruction that modifies them is encountered
and some operands have unknown values, the result is undetermined.
The outcome is thus picked up non-deterministically.

On the ARM9 processor, there are $4$ status bits.
A simple encoding would be to have $4$ boolean variables
to model the value of each bit.
As we let Player~2 choose the outcome, this corresponds to choosing
four values for Player~2: N (negative), Z (zero), V (overflow) and C
(carry). This could create $2^4=16$ different next states and thus as
many new potential branches in the game.  Most of the time, it is not
necessary to know the actual values of the $4$ status bits.  For
instance the result of a comparison instruction \textsf{cmp r0, r1}
with, say $\mathsf{r1}$ unknown, could be used later on only to check
wether $\mathsf{r0}=\mathsf{r1}$.  In this case the value of the
Z-status bit is required but the values of the other status bits are
irrelevant.

To reduce the number of branches (choices of Player~2) in the game, we
determine, for each instruction $\iota$ that sets a status bit, the
next instructions that depend on the result of $\iota$.  This can be
computed on the program $p$.  For each instruction $\iota$ that sets a
status bits, we let $\flag(\iota)$ be the set of predicates used after
$\iota$.  For instance in the example code of Listing~\ref{fib-as},
Fig.~\ref{fig-fibo} page~\pageref{fig-fibo}, the result of the
instruction \textsf{cmp r2, r0} line~4 is used at line~14, and the
only predicate needed is \textsf{gt} (\ie whether
$\mathsf{r2}>\mathsf{r0}$).  In the worst case we still need $4$
variables to encode the outcome of an instruction $\iota$ that sets
the status bits, but we reduce the choices of Player~2 to the
predicates in $\flag(\iota)$.  In the previous examples, instead of
having $16$ branches, there will be only $2$.



\noindent To model program $p_\bot$ in UPPAAL we need:
\begin{itemize}
\item an array, \texttt{val}, of $16$ variables for the registers of
  the ARM9 processor;
\item $4$ boolean variables for the status bits (we use \texttt{cmple},
    \texttt{cmplt}, \texttt{cmpls}, \texttt{cmpeq} ins\-tead of the actual status bits N, Z, V
  and C, but this is equivalent);
\item a \emph{stack} of size $K$ (the size of which has been
  determined in a previous stage).
\end{itemize}
Although the model-checker UPPAAL that we use is extremely efficient,
we have to be careful when encoding $p_\bot$: 
some information can be encoded using variables, but they will be part
of the \emph{state} of the network of TA we build, and will be encoded
in the BDD representation of each state.  Some information are not
dynamic but rather static (\eg the \emph{type} of an instruction
$\iota$, or the registers read/written $\iread(\iota)$ and
$\iwrite(\iota)$) and can be encoded using UPPAAL \emph{functions}.
This saves space as functions are not part of the encoding of a state.
Given a program $p_\bot$, we define the functions:
\begin{itemize}
\item $\setNZ : p \rightarrow \setB$ which, given an instruction
  $\iota \in p$, returns $\true$ if $\iota$ sets some status bits
  (comparison instructions \textsf{cmp,tst} and instructions with the
  ``s'' flag like \textsf{subs, adds} etc);
\item $\cmpU: p \times \calV_\bot \rightarrow \setB$ which returns
 $\true$ if the result of the instruction $\iota$ in
 state $v$ is unknown.
\end{itemize}
As a shorthand we write $\NDcmp(\iota,v) = \setNZ(\iota) \wedge
\cmpU(\iota,v)$ and this indicates whether instruction $\iota$, when
executed from state $v$, should be played by Player~2 (the status bits
should be set but an operand is unknown).

In addition to this, we define another function $\update: \calV_\bot
\rightarrow \calV_\bot$ which updates the values of the registers and
the status bits if required: this function encodes the semantics of
each instruction on the extended domain.

The result for the Fibonnaci program of Listing~\ref{fib-as-complete}
page~\pageref{fib-as-complete} are given in Listings~\ref{fib-C-setNZ}
and~\ref{fib-C-update}.
These listings call for some comments:
\begin{itemize}
\item Listing~\ref{fib-as-complete} contains the assembly code
  generated by \texttt{objdump} after compiling the C program with
  \texttt{gcc}; the instructions that set status bits have been
  annotated (\eg lien~4 \texttt{/ le /}) by the
  predicates that should be set by the instruction (\texttt{le} in
  this case for instructions at lines~4, 20 and~24).
\item Listing~\ref{fib-C-setNZ} contains the functions that determine
  whether the result of an instruction that sets the status bits is
  undetermined. \texttt{UNKNOWN} is a special value\footnote{We use an
    integer that is never used as an actual value in the content of
    any register.}.  For instance, if the value of \textsf{r2} is
  unknown when executing instruction (hexadecimal) $20$ (decimal
  $32$), \texttt{cmpU} returns $\true$ and \texttt{SetStatusB} as well.
\item Listing~\ref{fib-C-update} contains the updates of the registers
  in the extended domain.  The updates of an instruction are performed
  only if it is not abstracted away (\texttt{is\_abstracted} function,
  not given here, but we can assume for now it always returns
  $\false$.)  The instruction \textsf{cmp r2,r0} (UPPAAL translation
  lines~13 to~20) sets the \texttt{cmple} variable according to the
  values of \textsf{r2} and \textsf{r0}.  If at least one of the
  values of \textsf{r2} and \textsf{r0} is unknown, the value of
  \texttt{cmple} will be chosen right after the update step by
  Player~2, overriding the previous value.
  
  The instruction \textsf{cmp r2,r0} is \emph{unconditional}, and it
  has to be scheduled for execution.  This is carried out by function
  \texttt{SET(-,-,-)} which sets $3$ values (in the first stage of the
  pipeline, see section~\ref{sec-architecture}): the label of the instruction
  ($4$), the memory addresses referenced by the instruction ($-1$
  indicates no memory addresses), and wether the instruction is
  scheduled or not ($1$ in this case).

  For conditional instructions, \eg \textsf{movgt pc, lr}, (UPPAAL
  translation lines~24 to~37), if the function \texttt{gt()} returns
  $\true$, the instruction is not scheduled
  (\texttt{SET(20,-1,\textbf{0})}).  Function \texttt{gt()} returns
  the complement value of \texttt{cmple} that has been set by the
  comparison instruction (or Player~2 if some operands were unknown)
  before.

  The last parameter of \texttt{SET(-,-,-)} has no meaning for
  conditional branching instructions as they are always scheduled.  We
  use it to indicate whether the condition evaluates to $\true$ or
  $\false$. An example is instruction \texttt{ble 18} (UPPAAL
  translation lines~76 to~83 in listing~\ref{fib-C-update}).  If the
  condition (function \texttt{le()}) evaluates to $\true$ this
  parameter is $\true$ and $\false$ otherwise.  This information is
  used to simulate pipeline \emph{flushes} when a branch prediction is
  wrong.
\end{itemize}

\begin{center}
{\centering
\begin{minipage}{9cm}
\thicklines
\begin{lstlisting}[language=AssemblerARM9,numbers=none,caption={Complete Assembly Code},label={fib-as-complete}]
00000000 <fib>:
   0:	e3a02002 	mov	r2, #2	; 0x2
   4:	e1520000 	cmp	r2, r0                      / le / 
   8:	e1a0c000 	mov	ip, r0
   c:	e3a00001 	mov	r0, #1	; 0x1
  10:	e3a01000 	mov	r1, #0	; 0x0
  14:	c1a0f00e 	movgt	pc, lr
  18:	e2822001 	add	r2, r2, #1	; 0x1
  1c:	e1a03000 	mov	r3, r0
  20:	e352001e 	cmp	r2, #30	; 0x1e             / le /
  24:	d152000c 	cmple	r2, ip                     / le /
  28:	e0800001 	add	r0, r0, r1
  2c:	e1a01003 	mov	r1, r3
  30:	dafffff8 	ble	18 <fib+0x18>
  34:	e1a0f00e 	mov	pc, lr

00000038 <main>:
  38:	e1a0c00d 	mov	ip, sp
  3c:	e92dd810 	stmdb	sp!, {r4, fp, ip, lr, pc}
  40:	e3a0401e 	mov	r4, #30	; 0x1e
  44:	e24cb004 	sub	fp, ip, #4	; 0x4
  48:	e1a00004 	mov	r0, r4
  4c:	ebffffeb 	bl	0 <fib>
  50:	e1a00004 	mov	r0, r4
  54:	e91ba810 	ldmdb	fp, {r4, fp, sp, pc}
\end{lstlisting}
\end{minipage}
}
\end{center}

{\centering

\thicklines

\begin{lstlisting}[caption={C Code for \texttt{SetStatusB} and \texttt{cmpU}},label={fib-C-setNZ}]
/*  function to determine whether status bits should ne set */
bool SetStatusB(int i) { // i is the PC of instruction; function that tells whether status bits should be set
// comparisons for function fib
if (i==4) { // setting status bits for instruction cmp at 4 [0x4]
    return true ;
}
if (i==32) { // setting status bits for instruction cmp at 32 [0x20]
    return true ;
}
if (i==36) { // setting status bits for instruction cmp at 36 [0x24]
    return true ;
}
// comparisons for function main
return false ; 
}

/*  comparisons for instructions used in the program */
bool cmpU(int i) { 
/* comparisons for function fib starting 0 ending 52 */
if (i==4) return val[r2]==UNKNOWN||val[r0]==UNKNOWN; // [0x4]
if (i==32) return val[r2]==UNKNOWN; //  [0x20]
if (i==36) return val[r2]==UNKNOWN||val[ip]==UNKNOWN; //  [0x24]
/* comparisons for function main starting 56 ending 84 */
return false; // none if not found
} // end comp of instruction 

/*  setcmp for instructions used in the program */
void setcmp(int  i,bool n1,bool n2) { 
/* res_comp for function fib starting 0 ending 52 */
    if (i==4) { // instruction  cmp r2, r0 at 4 [0x4]
        cmple=n1;
    }
    if (i==32) { // instruction  cmp r2, #30 at 32 [0x20]
        cmple=n1;
    }
    if (i==36) { // instruction  cmple r2, ip at 36 [0x24]
        cmple=n1;
    }
/* res_comp for function main starting 56 ending 84 */
} // end setcmp of instruction 

bool NDcmp(int i) {
	return SetStatusB(i) && cmpU(i) ;
}

/*  setcmp for instructions used in the program */
void setcmp(int  i,bool n1,bool n2) { 
/* setcmp for function fib starting 0 ending 52 */
    if (i==4) { // instruction  cmp r2, r0 at 4 [0x4]
        cmple=n1;
    }
    if (i==32) { // instruction  cmp r2, #30 at 32 [0x20]
        cmple=n1;
    }
    if (i==36) { // instruction  cmple r2, ip at 36 [0x24]
        cmple=n1;
    }

/* res_comp for function main starting 56 ending 84 */

} // end setcmp of instruction 

\end{lstlisting}
}

{\centering\thicklines
\begin{lstlisting}[caption={C Program},label={fib-C-update}]
void update() { // update function 
int nextpc,nextfp,tmp;
/*
 updates for function fib starting 0 ending 52
*/
if (val[pc]==0) { // Instruction mov r2, #2 at 0x0
    nextpc=val[pc]+4;
    if (!is_abstracted(val[pc])) { // effect of instruction is null if abstracted
        val[r2]=(2);
    }
    SET(0,-1,1); // instruction scheduled is 0, no memory access and scheduled
} // end mov at 0x0
if (val[pc]==4) { // Instruction cmp r2, r0 at 0x4
    nextpc=val[pc]+4;
    if (!is_abstracted(val[pc])) { // effect of instruction is null if abstracted
        // Should set the Z and N and C bits
        if ((val[r2]-(val[r0]))<=0) cmple=1 ; else cmple=0; 
    }
    SET(4,-1,1); // instruction scheduled is 4, no memory access and scheduled
} // end cmp at 0x4

...

if (val[pc]==20) { // Instruction movgt pc, lr at 0x14
    nextpc=val[pc]+4;
    if (gt()) {
        if (!is_abstracted(val[pc])) { // effect of instruction is null if abstracted
            if (val[lr]==UNKNOWN) {
                val[pc]=UNKNOWN;
            }
            else {
                nextpc=(val[lr]);
            }
        }
        SET(20,-1,1); // instruction scheduled is 20, no memory access and scheduled
    }
    else SET(20,-1,0) ; // instruction not scheduled, no mem access
} // end movgt at 0x14
if (val[pc]==24) { // Instruction add r2, r2, #1 at 0x18
    nextpc=val[pc]+4;
    if (!is_abstracted(val[pc])) { // effect of instruction is null if abstracted
        if (val[r2]==UNKNOWN) {
            val[r2]=UNKNOWN;
        }
        else {
            val[r2]=(val[r2]+1);
        }
    }
    SET(24,-1,1); // instruction scheduled is 24, no memory access and scheduled
} // end add at 0x18

...

if (val[pc]==32) { // Instruction cmp r2, #30 at 0x20
    nextpc=val[pc]+4;
    if (!is_abstracted(val[pc])) { // effect of instruction is null if abstracted
        // Should set the Z and N and C bits
        if ((val[r2]-(30))<=0) cmple=1 ; else cmple=0; 
    }
    SET(32,-1,1); // instruction scheduled is 32, no memory access and scheduled
} // end cmp at 0x20
if (val[pc]==36) { // Instruction cmple r2, ip at 0x24
    nextpc=val[pc]+4;
    if (le()) {
        if (!is_abstracted(val[pc])) { // effect of instruction is null if abstracted
            // Should set the Z and N and C bits
            if ((val[r2]-(val[ip]))<=0) cmple=1 ; else cmple=0; 
        }
        SET(36,-1,1); // instruction scheduled is 36, no memory access and scheduled
    }
    else SET(36,-1,0) ; // instruction not scheduled, no mem access
} // end cmple at 0x24

...

if (val[pc]==48 && (!le())) { // Instruction ble 18,  at 0x30
    nextpc=val[pc]+4;
    SET(48,-1,0) ; // instruction scheduled, no mem access, no branching
} // end ble at 0x30 [cond false]
if (val[pc]==48 && le()) { // Instruction ble 18,  at 0x30
    nextpc=24; // to 0x18
    SET(48,-1,1) ; // instruction scheduled, no mem access, branching
} // end ble at 0x30 [cond true]
if (val[pc]==52) { // Instruction mov pc, lr at 0x34
    nextpc=val[pc]+4;
    if (!is_abstracted(val[pc])) { // effect of instruction is null if abstracted
        if (val[lr]==UNKNOWN) {
            val[pc]=UNKNOWN;
        }
        else {
            nextpc=(val[lr]);
        }
    }
    SET(52,-1,1); // instruction scheduled is 52, no memory access and scheduled
} // end mov at 0x34

/*
 end of updates for function fib
*/

/*
 updates for function main starting 56 ending 84
*/
if (val[pc]==56) { // Instruction mov ip, sp at 0x38
    nextpc=val[pc]+4;
    if (!is_abstracted(val[pc])) { // effect of instruction is null if abstracted
        if (val[sp]==UNKNOWN) {
            val[ip]=UNKNOWN;
        }
        else {
            val[ip]=(val[sp]);
        }
    }
    SET(56,-1,1); // instruction scheduled is 56, no memory access and scheduled
} // end mov at 0x38
if (val[pc]==60) { // Instruction stmdb sp!,{r4,fp,ip,lr,pc,} at 0x3c
    nextpc=val[pc]+4;
    // push should first decrease val[pc] and then store in stack(val[pc])
    push(val[pc]);
    push(val[lr]);
    push(val[ip]);
    push(val[fp]);
    push(val[r4]);
    SET(60,-1,1); // instruction scheduled is 60, no memory access
} // end stmdb at 0x3c

...

if (val[pc]==76) { // Instruction bl 0,  (unconditional) at 0x4c
    nextpc=0; // to 0x0
    val[lr]=80;
    SET(76,-1,1) ; // instruction scheduled, no mem access, branching
} // end bl at 0x4c
if (val[pc]==80) { // Instruction mov r0, r4 at 0x50
    nextpc=val[pc]+4;
    if (!is_abstracted(val[pc])) { // effect of instruction is null if abstracted
        if (val[r4]==UNKNOWN) {
            val[r0]=UNKNOWN;
        }
        else {
            val[r0]=(val[r4]);
        }
    }
    SET(80,-1,1); // instruction scheduled is 80, no memory access and scheduled
} // end mov at 0x50
if (val[pc]==84) { // Instruction ldmdb fp,{r4,fp,sp,pc,} at 0x54
    nextpc=val[pc]+4;
    nextpc=stack(val[fp]-4);
    val[sp]=stack(val[fp]-8);
    nextfp=stack(val[fp]-12);
    val[r4]=stack(val[fp]-16);
    val[fp]=nextfp;
    SET(84,-1,1); // instruction scheduled is 84, no memory access
} // end ldmdb at 0x54

/*
 end of updates for function main
*/

val[pc]=nextpc;
} // end update 
\end{lstlisting}
}

\begin{figure}[hbtp]
  \centering
  \includegraphics[scale=0.5]{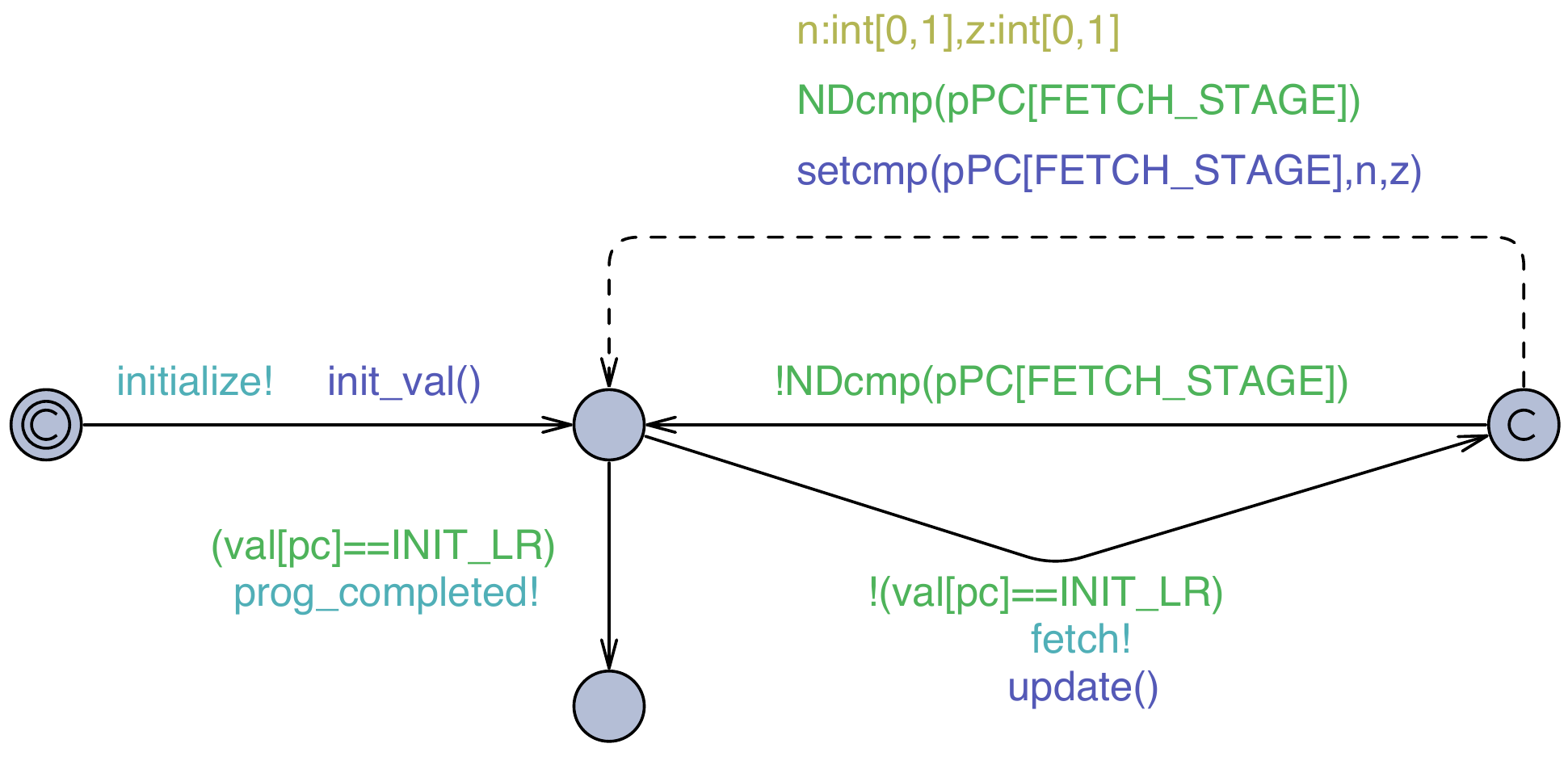}
  \caption{Generic Automaton \emph{Prog} to Simulate a Program}
  \label{fig-prog-auto}
\end{figure}

The generic automaton to simulate a program $p_\unk$ is given in
Fig.~\ref{fig-prog-auto}.  We assume that the main function of the
program $p_\unk$ is called by another program and a particular value
\texttt{INIT\_LR} gives the return point.  The automaton \emph{Prog}
performs some initialization (\texttt{init\_val()}) and then computes
the next state until the end of the program is reached: this is when
the value of the \textbf{pc} register is equal to the return point
\texttt{INIT\_LR} (guard \texttt{val[pc]=INIT\_LR}).  To simulate each
instruction, the automaton \emph{Prog} performs the following steps:
\begin{enumerate}
\item feed the current instruction $\iota$ to the first stage of the
  pipeline when it is empty (to do so it has to synchronize with the
  first stage of the pipeline, on the \texttt{fetch!} channel) and
  compute the next state (\texttt{update()} function).  This also sets
  the next value of register \textbf{pc}. The result of
  \texttt{update()} is that the number of the current instruction is
  stored into the variable \texttt{pPC[FETCH\_STAGE]} where
  \texttt{FETCH\_STAGE} is the number of the first stage of the
  pipeline ($0$);
\item if the instruction $\iota$ in \texttt{pPC[FETCH\_STAGE]} is an
  undetermined comparison (\texttt{NDcmp(pPC[FETCH\_STAGE])} evaluates
  to $\true$), the upper dashed transition is taken: Player~2 chooses
  two values $n$ and $z$ and the predicates that must be set
  (\texttt{cmple}, \texttt{cmplt}, etc) are set by \texttt{setcmp}
  (Listing~\ref{fib-C-setNZ}).  If $\iota$ does not set any flag or
  the outcome is determined by the current state (the operands are all
  known), the middle transtion is taken (Player~2 does not have to
  play).
\end{enumerate}

\section{Model of  the Hardware}
\label{sec-architecture}
In this section we give a UPPAAL model for the architecture of the
pipelined processor ARM9 and for the caches.

\subsection{Model of the Pipeline}
Each stage of the pipeline contains an instruction (and some other
information).  The information for each stage of the pipeline are
stored in arrays: \texttt{pPC[k]} gives the number of the instruction
in stage $k$; \texttt{Todo[k]} is a boolean value and indicates
whether the instruction \texttt{pPC[k]} is scheduled (some
instructions are conditional and are skipped); \texttt{dataAdr[k]}
contains the address\footnote{For multiple loads and stores, this
  should be a range of addresses; this information is used only for
  determining whether a stall should occur in the pipeline.  For
  multiple loads and stores, we force a stall in a pipeline until the
  end of the multiple loads/stores instruction. This is a safe
  encoding as the ARM9 does not exhibit timing anomalies.} of the
memory cell referenced by instruction \texttt{pPC[k]} ($-1$ if none).
There are $5$ stages in the pipeline of the ARM9:
\begin{itemize}
\item stage~1: this is the \emph{fetch} stage. It fetches the next
  instruction (pointed to by the \textbf{pc} register) from the cache
  (or main memory) and this instruction becomes the current instruction of
  stage~1;
\item stage~2: \emph{decode} stage. Decodes the instruction in stage~2;
\item stage~3: \emph{execute} stage. Carries out the computation
  (addition, comparisons, etc) of the instruction in stage~3;
\item stage~4: \emph{memory} stage.  Carries out the transfers (from
  registers to main memory or main memory to registers) of the instruction
  in stage~4;
\item stage~5: \emph{writeback} stage. Writes the value of registers
  that are (``writeback'') operands of the instruction in stage~5. 
\end{itemize}
An instruction $\iota$ enters the pipeline at stage~1.  It is
transfered from stage $i$ to $i+1$ as soon as possible.  When it exits
stage~5, it is completed.  The execution of a program is completed
when its last instruction is completed.

\paragraph{\bfseries Pipeline Stalls.}
The goal of \emph{pipelining} is to split the execution of an
instruction into different simple steps.  The idea being that each
step can be carried out concurrently for different instructions: while
stage~1 fetches the next instruction $\iota_k$, stage~2 decodes
instruction $\iota_{k-1}$, etc.  It may happen that the simple steps
of some sequences of instructions cannot be carried out concurrently.
A \emph{pipeline stall} is a situation when one stage $i$ of the
pipeline cannot perform its computation because it has to wait for
another stage $j>i$ to complete its computation.  An example is when
the execution of an instruction at stage~3 (execute) has an operand
which is set in stage~4 (memory).

\noindent\begin{minipage}[T]{0.6\linewidth}
  The sequence of instructions of lines~0 and~4 will result in a
  pipeline stall at stage~3 for instruction $4$: when instruction $4$
  ($\mathsf{r2}:=\mathsf{r0}-\mathsf{r1}$) is ready to execute at
  stage~3, it has to wait for instruction $0$ to complete (at stage~4)
  because instruction $0$ loads the value of memory cell $\mathsf{r1}$
  into \textsf{r0}. 
\end{minipage}\hfill
\begin{minipage}[T]{0.3\linewidth}
\centering
\thicklines
\begin{lstlisting}[language=AssemblerARM9,numbers=none,caption={Stalls},label={list-stall}]
   0:    ldr r1, [r0]
   4:    sub r2, r0, r1
   8:    ... 
   c:    ldm r13, {r1,r2,r3}
  10:    add r4, r3, #1
  14:    ...
\end{lstlisting}
\end{minipage}\hspace{.05\linewidth}

Thus instruction $4$ stalls for one cycle\footnote{We assume that the
  content of memory cell was in the cache and it takes one cycle to be
  fetched.} at stage~$3$.  The situation for instructions~c and~10 can
even result in more than one cycle delay.  The \textsf{ldm}
isntruction (line~c) is a \emph{multiple load} instruction.  It loads
the registers \textsf{r1}, \textsf{r2} and \textsf{r3} with the
contents of memory cells pointed to by \textsf{r13}.  Stage~4 performs
the loads, but only one per cycle.  Thus instruction~10 stalls for 3
cycles at stage~3.

A pipeline stall may occur depending on: ($i$) the type of the
instruction at stage~3, and the type of the instruction at stages~4
and~5; ($ii$) the registers (and memory addresses) used by the
instructions at the corresponding stages.

\paragraph{\bfseries Branch Prediction.}
When a conditional branch instruction enters the pipeline, the next
instruction to flow in is determined by the truth value of the
condition. This value might not yet be available when the branch
instruction is in the first stage of the pipeline.  If the condition
is determined by the value of a variable which is not in the cache, it
might take a few cycles before the result becomes available.  In this
case, we should \emph{stall} until the outcome of the comparison is
computed. This might however be inefficient.

Some heuristics can be applied to guess the most plausible next
instruction after a conditional branching.  After the
\emph{prediction}, the chosen instruction flows in the pipeline.  If
the guess was right the result is shortest execution time for this
part of the program.  If the guess was wrong, the computations of the
mistakenly taken branch have to be undone, and the pipeline flushed
which results in a longer execution time.  We do not discuss here the
choice of a good heuristics, but there are a few options that gve good
results on \emph{average}. 

In our model we follow~\cite{metamoc-2009} and model the heuristics
for branch prediction by: in a conditional branch, a branch is never
taken (other heuristics can be accommodated for in our model).

\paragraph{\bfseries UPPAAL Pipeline Model.}
The timed automata models we introduce are close to the ones proposed
in~\cite{metamoc-2009}. However there are some differences as we do
not have the same model for the program.

\begin{figure}[hbtp]
  \centering
  \includegraphics[scale=0.6]{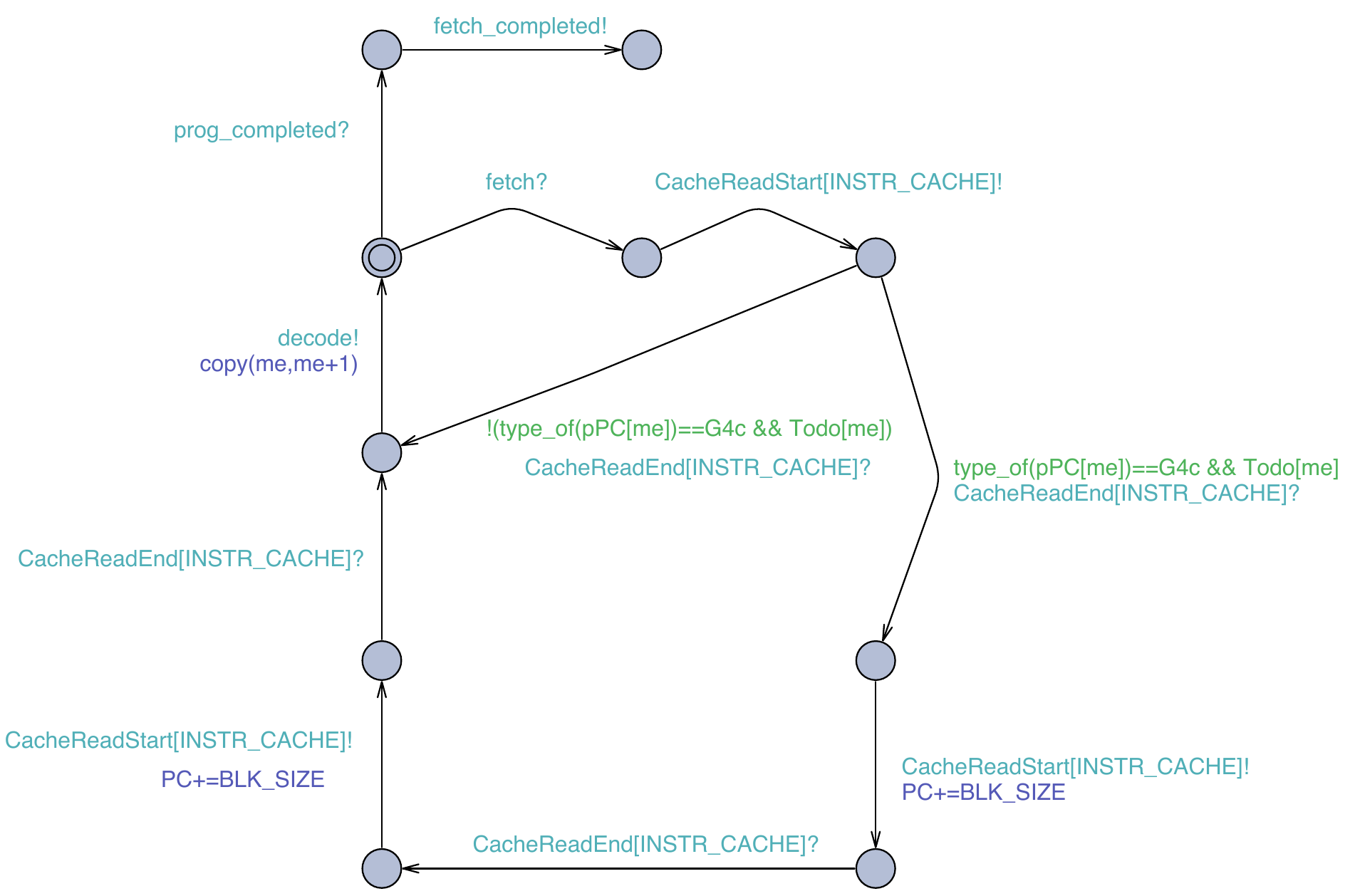}
  \caption{Timed Automata Model of the ARM9 Pipeline}
\label{fig-fetch-stage}
\end{figure}

\begin{figure}[hbtp]
  \centering
  \includegraphics[width=0.544\linewidth]{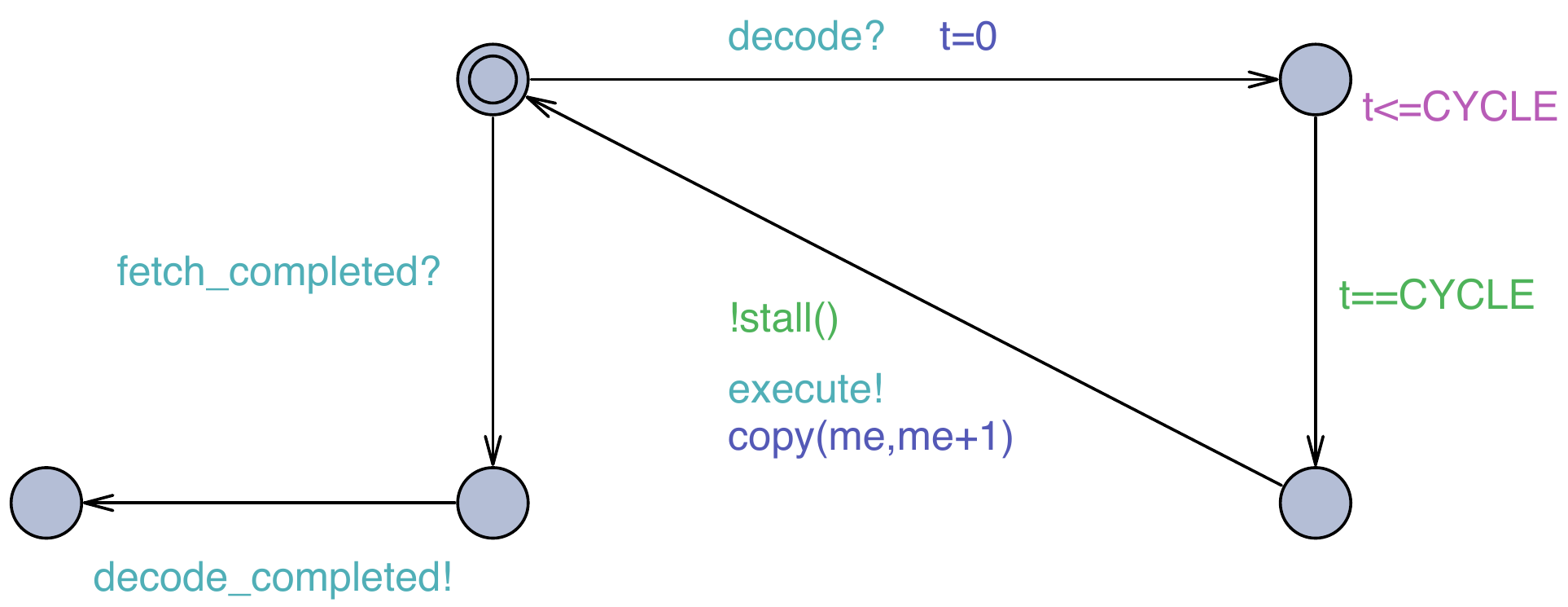}
  \bigskip

  \includegraphics[scale=0.6]{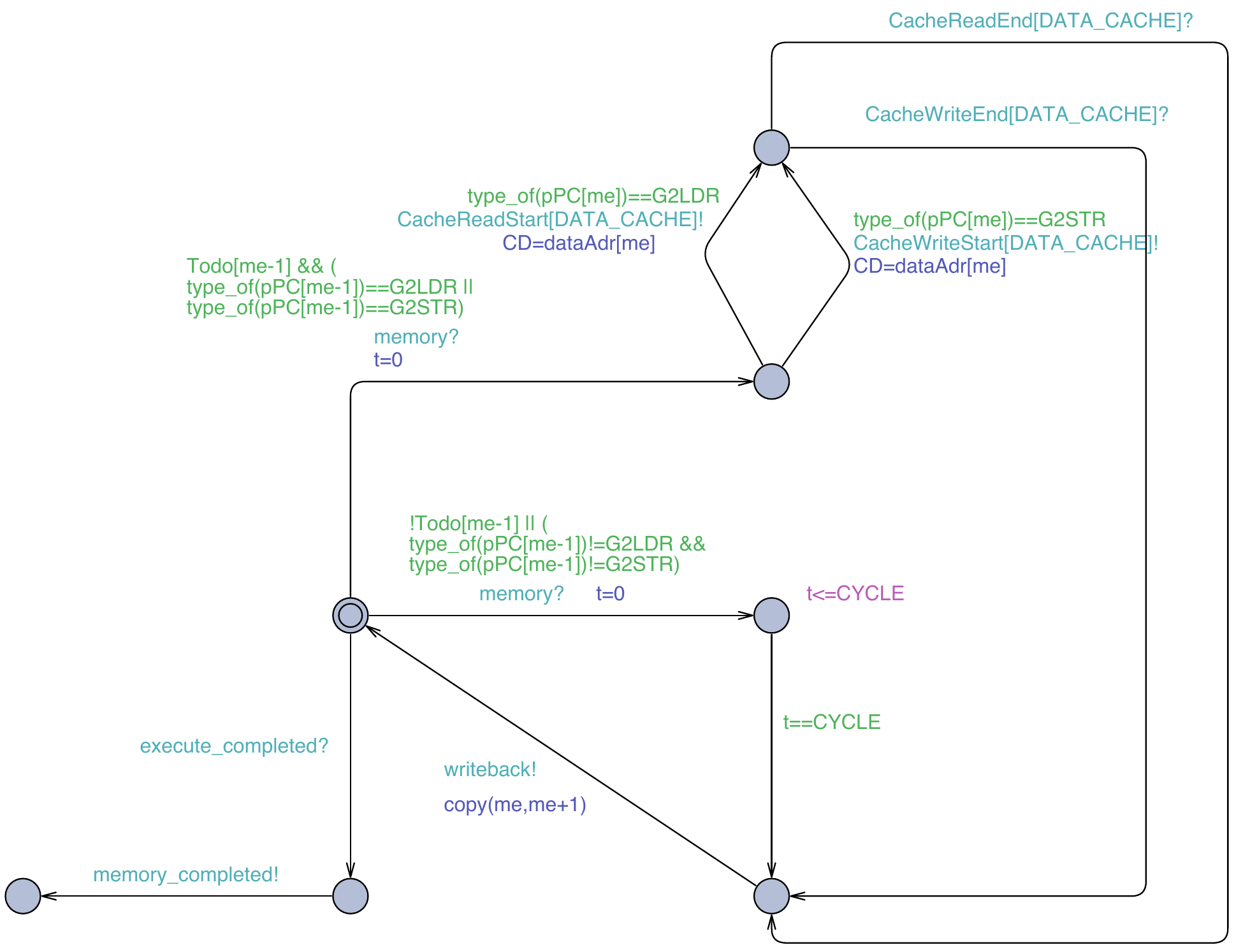}
  \bigskip

  \includegraphics[width=0.54\linewidth]{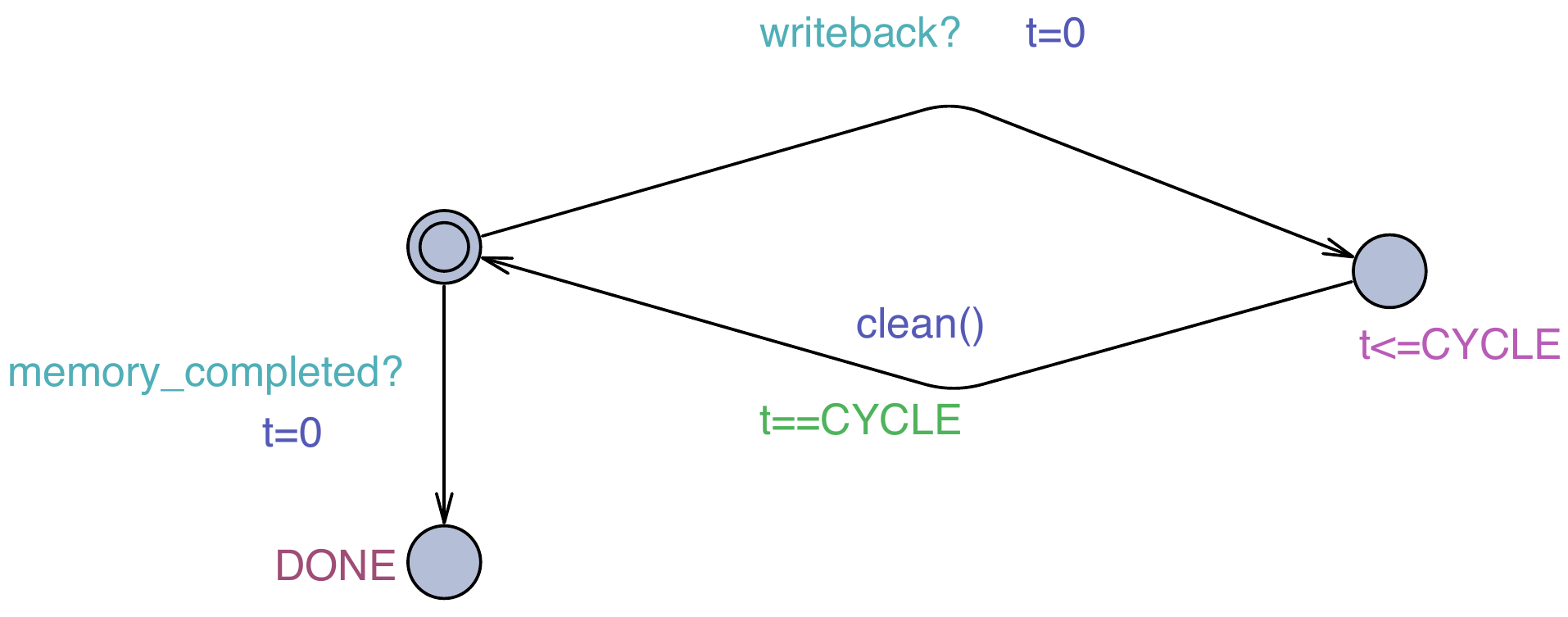}
  \caption{Timed Automata Model of the ARM9 Pipeline}
\label{fig-other-stage}
\end{figure}

The timed automata for each stage (ARM9, 5 stages) are depicted on
Fig.~\ref{fig-fetch-stage} and Fig.~\ref{fig-other-stage}.  The stage
modelled by each automaton can be infered by the synchronization
channel from the initial state (\eg \texttt{decode?}).
The first stage of the pipeline is of particular importance as it
models the case of a wrong guess in an branch prediction.  The
automaton of Fig.~\ref{fig-fetch-stage} models the following
behaviour:
\begin{enumerate}
\item the automaton accepts a \texttt{fetch?} synchronization
  when it is idle;
\item after accepting an instruction (\texttt{fetch?} synchronizes
  with \texttt{fetch!} in the automaton \emph{Prog} of
  Fig.~\ref{fig-prog-auto}), it actually fetches the instruction from main
  memory via the \emph{instruction cache}
  (\texttt{CacheReadStart[INSTR\_CACHE]!}, where \texttt{INSTR\_CACHE}
  is the ID or the instruction cache);
\item when the instruction has been read from the cache or main
  memory, there are two options:
  \begin{enumerate}
  \item the instruction $\iota$ to be processed is a \emph{conditional
      branch} (condition \texttt{type\_of(pPC[me])==G4c}) and the
    variable \texttt{Todo[me]} indicates whether the condition was
    evaluated to $\true$ or $\false$.  In case it is a conditional
    branch and the condition was $\true$, we simulate two
    ``instruction read from the cache'' steps: indeed our branch
    prediction algorithm is ``never branch'' and thus if it happened
    that we had to branch, we should simulate a pipeline
    \texttt{flush}. As we do not execute the instructions in the
    pipeline (but rather when we feed the first stage of the
    pipeline), this can be modelled by reading the next two
    instructions (the ``never branch'' prediction) without executing
    them, and then resuming the simulation from the target address of
    the branch instruction.
  \item the instruction to be processed is not a conditional branching
    or the condition was evaluated to $\false$; in this case the
    prediction was right and nothing has to be undone.
  \end{enumerate}
  After an instruction has been fetched in the \texttt{fetch} stage,
  it is fed to the next stage of the pipeline. This is modelled by the
  \texttt{decode!} synchronization and the \texttt{copy(me,me+1)}
  transition. \texttt{copy(me,me+1)} copies the information in
  \texttt{pPC[me]}, \texttt{Todo[me]} and \texttt{dataAdr[me]} to the
  next stage \texttt{me+1}.
\end{enumerate}

The memory stage automaton is a bit more involved than the others as
it has to take into account different options: if the instruction is a
memory transfer (\texttt{type\_of(pPC[me-1])==G2LDR} or
\texttt{type\_of(pPC[me-1])==G2STR}) and is scheduled
(\texttt{Todo[me-1]} is $\true$) a synchronization with the data cache
is requested.

\smallskip

The type of the instructions is given by a UPPAAL function
\texttt{type\_of}. The duration is also given by a function
\texttt{dur()} (used in the execute stage).

\subsection{Model of the Caches}
A \emph{cache} is a fast memory device.  It is characterized by its
size $K$ (usually in Kbytes), the length of a cache \emph{line} ($B$
in Bytes) and the number of cache lines $L = \frac{K}{B}$.

The main memory $\calM$ of a computer is divided into blocks equal to
the length of the cache line.
We let $\calM=\{m_0,m_1,\cdots,m_n\}$.

\noindent The \emph{associativity} of a cache determines where a
memory block can reside.
\begin{itemize}
\item \emph{fully associative}: a block can be in any line;
\item \emph{direct mapped}: a block can be in one line;
\item \emph{$j$-way}: a block can be in $j$ different lines; in this
  case the cache is partitionned into $\frac{L}{j}$ different sets.
  Fully and direct mapped are particular instances of $j$-way caches.
  The partition induced by the $j$-way cache is denoted
  $\calP=\{P_1,\cdots,P_{\frac{L}{j}}\}$.
\end{itemize}
The set of lines a memory can reside in is given by a mapping $\kappa
: \calM \rightarrow \calP$.

\noindent The replacement policy determines which block to eject from memory
when the cache is full. The most common policies are:
\begin{itemize}
\item LRU: least recently used;
\item FIFO: first-in first-out;
\item alternate and mixed and even random are permitted but not easily
  predictable.
\end{itemize}
\begin{figure}[hbtp]
  \centering
 \includegraphics[width=1\linewidth]{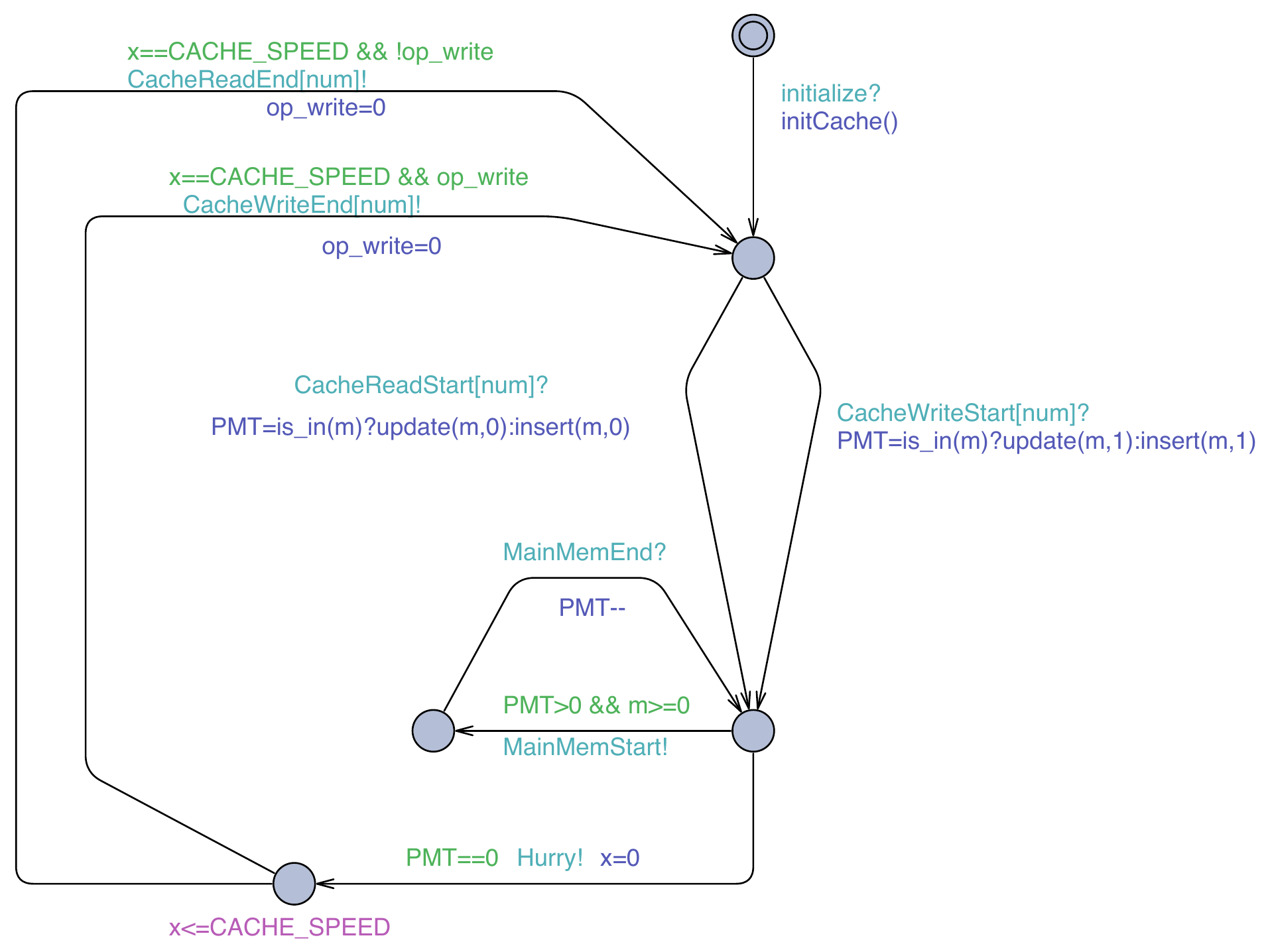}
  \bigskip

  \includegraphics[scale=0.3]{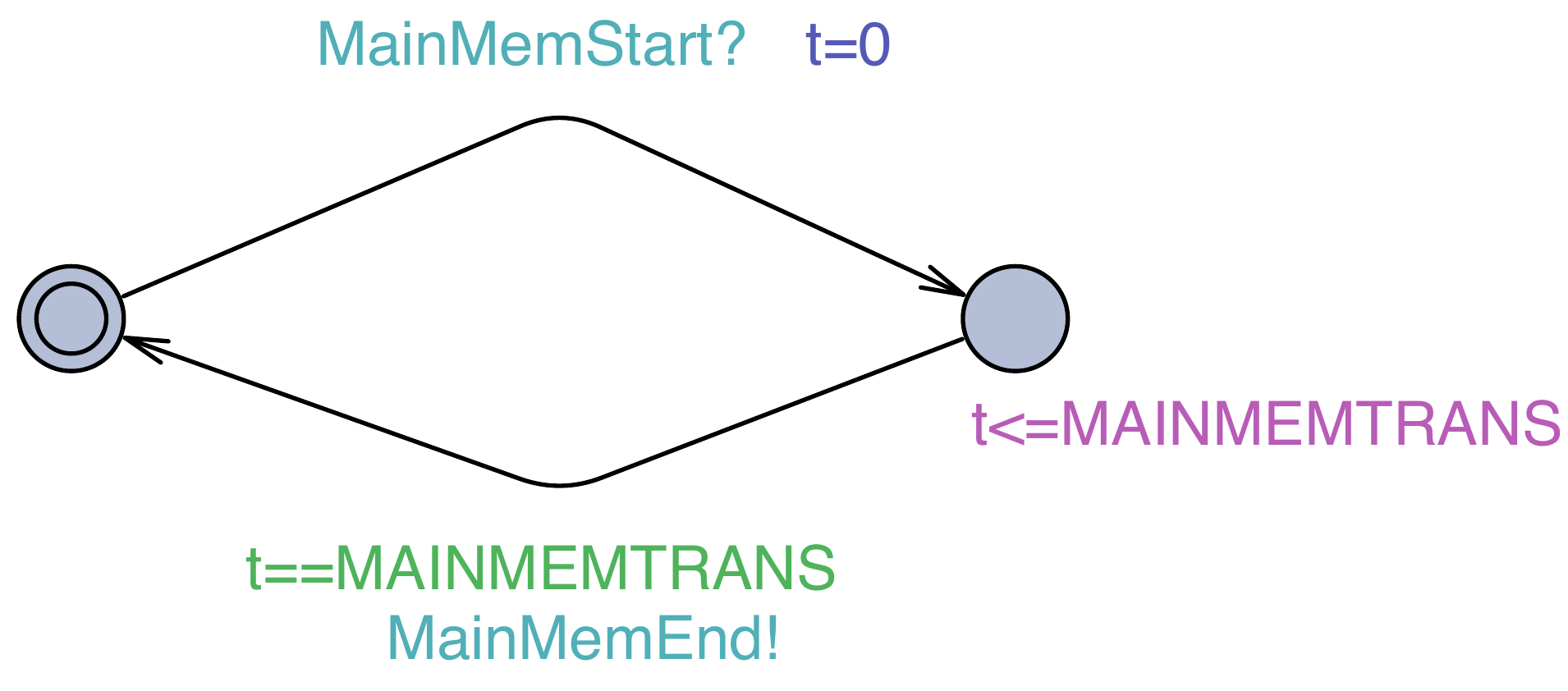}

  \caption{Timed Automata Model for the Caches}
\label{fig-cache}
\end{figure}
Handling \emph{writing} requests is also a distinctive feature of a
cache.
\begin{itemize}
  \item handling write \emph{hits}: 
    \begin{itemize}
    \item \textbf{write trough}: write cache and memory
    \item \textbf{write back}: write cache; need for a dirty bit whihc is taken
      care of when ejecting a line from the cache;
    \end{itemize}
  \item handling write \textsf{misses}:
    \begin{itemize}
    \item \textbf{write allocate}: write memory and fetch into cache;
    \item \textbf{write no allocate}: write memory (no fetch).
    \end{itemize}
\end{itemize}
In this paper we model a cache with FIFO replacement policy and assume
write allocate on a write/miss.

\paragraph{\bfseries UPPAAL Cache Model.}

The automaton modeling the behaviour of the cache (together with the
model iof the main memory automaton) is given in Fig.~\ref{fig-cache}.

After performing some initializations (\texttt{initCache()}, setting
the initial state of the cache), it accepts either write or read
requests.  Depending on the request, and wether a cache line is dirty
or not, a number of memory transactions (\texttt{PMT}) are needed to
fetch the content of memory cell \texttt{m}.  Each such transaction is
performed one after the other.  When it is completed the transfer from
the cache to the register of the processor takes place and require
\texttt{CACHE\_SPEED} time units.

\section{Tool Chain and Case Studies}
\label{sec-tool-chain}
We have applied the previous framework to a number
of benchmarks from \malar.

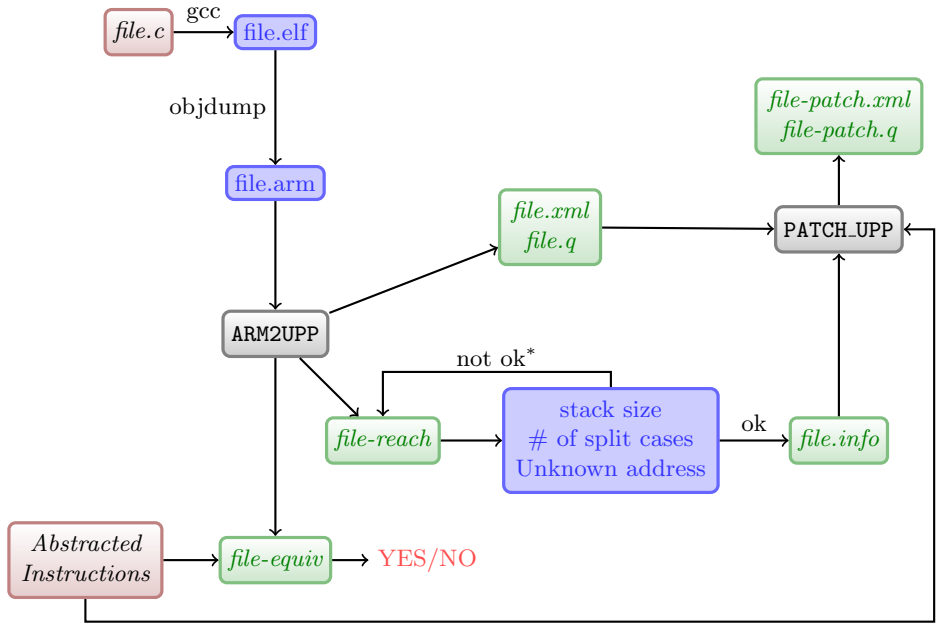
\begin{figure}[hbtp]
\centering
\begin{tikzpicture}[
  input/.style={ 
    rectangle, 
rounded corners=0.9mm,
    minimum size=6mm, 
    very thick, 
    draw=red!50!black!50, 
    top color=white, 
    bottom color=red!50!black!20, 
    font=\itshape 
},
  output/.style={ 
    color=green!50!black,
    rectangle, 
    minimum size=6mm, 
    rounded corners=0.9mm,
    very thick, 
    draw=green!50!black!50, 
    top color=white, 
    bottom color=green!50!black!20, 
    font=\itshape 
},
inter/.style={color=blue!80,rectangle,rounded corners=1mm,very thick,fill=blue!20,draw=blue!60},
mycomp/.style={ 
  rectangle,minimum size=6mm,rounded corners=1mm, 
  very thick,draw=black!50, top color=white,bottom color=black!20,
  font=\ttfamily}, node distance=2cm, thick ]

\node (c-file) [input] {file.c};
\node (elf-file) [inter, right=of c-file,xshift=-.2cm] {file.elf};
\path[->] (c-file) edge[] node {gcc} (elf-file) ;
\node (arm-file) [inter, below=of elf-file] {file.arm};
\path[->] (elf-file) edge[swap] node {objdump} (arm-file) ;

\node (comp) [mycomp,below=of arm-file] {ARM2UPP};
\path[->] (arm-file) edge[swap]  (comp) ;

\node (xml) [output,above right=of comp,xshift=2.2cm] {
  \begin{tabular}[h]{c}
file.xml \\ file.q
\end{tabular}
};
\node (reach) [output,below right=of comp] {file-reach};
\node (equiv) [output,below=of comp,yshift=-1cm] {file-equiv};

\path[->] (comp) edge[]  (xml);
\path[->] (comp) edge (reach);
\path[->] (comp) edge (equiv);

\node (abstract) [input,left=of equiv,xshift=-.5cm] {
  \begin{tabular}[h]{c}
    Abstracted \\ Instructions
\end{tabular}
}; 
\node (res-equiv) [right=of equiv,color=red!70] {YES/NO};
\path[->] (abstract) edge (equiv);
\path[->] (equiv) edge (res-equiv);

\node (reach-out) [inter,right=of reach,xshift=1cm] {
  \begin{tabular}[h]{c}
    stack size \\
    \# of split cases \\
    Unknown address 
  \end{tabular}
} ;
\node (ok) [output,right=of reach-out,xshift=1cm] {file.info};
\path[->] (reach) edge (reach-out) (reach-out) edge node {ok} (ok);
\draw[->] ($ (reach-out.north) $) -- ($ (reach-out.north) + (0,2mm) $)  node[xshift=-1.5cm,yshift=2mm] {not ok$^*$} -| 
    ($ (reach.north)  $) ;

\node (patch) [mycomp,above=of ok,yshift=.8cm] {PATCH\_UPP};
\path[->] (xml) edge (patch) ;
\path[->] (ok) edge (patch);

\node (file-patch) [output,above=of patch,yshift=-.5cm] {
  \begin{tabular}[h]{c}
file-patch.xml \\ file-patch.q
\end{tabular}
};
\path[->] (patch) edge (file-patch);

\draw[->] ($ (abstract.south) $) |- ($ (abstract.south) - (0,3mm) $)  -| 
    ($ (patch.east) + (.4cm,0mm)  $) -- ($ (patch.east) $) ;





\end{tikzpicture}

\caption{Tool Chain Overview}
\label{fig-tool-chain}
\end{figure}

\paragraph{\bfseries Tool Chain.}
The tool chain to compute WCET is depicted on
Fig.~\ref{fig-tool-chain}.  The component we have developed are
\texttt{ARM2UPP} and \texttt{PATCH\_UPP}:
\begin{itemize}
\item \texttt{ARM2UPP} takes as input a program in assembly
  (\texttt{file.arm}) that has been annotated with the comparisons
  operators for each instruction that sets a status bit.
  It generates four files:
  \begin{itemize}
  \item \texttt{file.\{xml,q\}} that contain respectively the UPPAAL
    network automata (and functions like \texttt{update()} etc)
    modeling the execution of the program on the architecture of the ARM9
    and the UPPAAL queries to compute/check the WCET;
  \item \texttt{file-reach} is an executable obtained by compiling
    \texttt{file-reach.cpp}; this latter file is a C++ program that
    simulates the program in \texttt{file.arm}. \texttt{file-reach}
    always terminates. However, early termination can be forced by
    passing some parameters (maximal number of states, maximal number
    of split cases).  In case the number of split cases is too large
    (\eg $2^{50}$ for Bubble Sort), it is possible to add some
    information in the file \texttt{file-reach.cpp} like constraints
    on the outcome of an unknown comparison. This step may be iterated
    several times. When it is completed the file \texttt{file.info}
    contains some useful information (like maximal stack size, etc).
  \item \texttt{file-equiv} is an executable obtained by compiling
    \texttt{file-equiv.cpp}; this program checks whether an
    abstraction mapping (which is given by a function) is valid or
    not (implements the algorithm of section~\ref{sec-concrete-abs}.).
  \end{itemize}
\item \texttt{PATCH\_UPP} modifies some constants in \texttt{file.xml}
  to incorporate the information from \texttt{file.info} (like stack
  size) and can also include the function of abstracted instructions
  (if it has been declared valid).
\end{itemize}

\paragraph{\bfseries UPPAAL-TiGA Queries.}
In order to compute the WCET of a program, we can check wether
the program always terminates within $k$ time units.  This can be
computed using a binary search with UPPAAL.  The drawback of this
check is that some deadlock may occur in the system, yielding
a biased value of the WCET.

An alternative way of computing the WCET is check a \emph{control}
property: ``Can Player~1 enforce termination of the program and if
yes, what is the best duration he can guarantee?''  This optimal time
reachability control objective can be checked in one query
(see~\cite{cassez-concur-05}) with UPPAAL-TiGA, provided we
know an upper of the WCET.  This can be roughly over-estimated on the
program (we have not implemented this part yet).
Optimal reachability of a location \texttt{l} is then specified by the
control objective:
\begin{center}
\begin{verbatim}
control(#n,0) : A [ true U  l ]
\end{verbatim}
\end{center}

if \texttt{\#n} is a rough upper bound of the WCET\footnote{If
  \texttt{\#n} is not large enough, UPPAAL-TiGA result will be ``not
  controllable''.}.

Program termination in the UPPAAL model happens when the location
\texttt{DONE} is reached in the \texttt{writeBackStage} automaton
(last stage of the pipeline). Thus the control property we check is:
\begin{center}
\begin{verbatim}
control(#n,0) : A [ true U  WriteBackStage.DONE ]
\end{verbatim}
\end{center}

\paragraph{\bfseries case Studies \& Results}
We have applied the framexork described in Fig.~\ref{fig-tool-chain}
to a number of benchmark programs from \malar.  We could not analyse
the full set of programs because of the current limitations of our
tools:
\begin{itemize}
\item floating point operations are not supported yet;
\item a few operators (\eg \texttt{ror}) of the ARM9 assembly language
  are not supported yet.
\end{itemize}
There are not many published results about the actual WCET of the
benchmarks (or when there are, the hardware parameters, cache speed,
etc are not given).  To evaluate the relevance of our method, we
compare our results to the ones obtained with the METAMOC
method~\cite{metamoc-2009}.

There are $15$ programs that can be analysed by METAMOC using a
concrete instruction cache and an ``always miss'' data cache.  Only
$7$ of the $15$ programs can be analysed with both a concrete
instruction and data cache.  Using our encoding and tool chain, we
could analyse $13$ out of these $15$ programs (two of them contains
unsupported operations) with concrete caches.  Moroever, the
time/space needed to compute the results is very small compared to the
resources used in METAMOC (32GB RAM computer).
Table~\ref{tab-results} give the values of WCET for each program, and
the time for UPPAAL-TiGA to compute the result.  The time needed to
compute the intermediary files is negligible.  The timing
specification of the caches are: \texttt{CACHE\_SPEED=1} (processor)
cycle is the same as the processor speed, and a memory transaction
takes $10$ processor cycles.
The UPPAAL files are available from \url{http://www.irccyn.fr/franck/wcet}.

\paragraph{\bfseries Energy/Power Consumption Optimization.}
The last column of Table~\ref{tab-results} gives the percentage of
time the processor can run at a slower clock rate ($1/4$th of its
fastest speed) without any impact on the WCET: this is due to the
initial transient phase of the execution of a program where
instructions are loaded into the cache.  For some small programs the
result is impressive (22\% for \texttt{janne-complex}).  To do this we
just add a automaton to the network that switches the rate from $4$ to
$1$ after a certain amount of time.  Another interesting and easy
computation that can be done, is to fix the time the processor runs at
a slower rate (in the initial phase) and compute the optimal time to
reach the end the program (which is the WCET) under this constraint.

\begin{table}[bhbtp]
  \centering
  \begin{tabular}{||l||c|c|c|c|c|c||}\hline\hline
    ~~\textbf{Program}~~ &
    \begin{tabular}[h]{c}
loc$^\dagger$
\end{tabular}
&  ~
\begin{tabular}[h]{c}
$N^\ddagger$
\end{tabular}
~ &  ~
    \begin{tabular}[h]{c}
UPPAAL-TiGA \\ time/space
\end{tabular}
~ &  ~WCET~ & ~~~Abs$^\star$~~~ & ~~~~\begin{tabular}[h]{c}
Low \\ Power
\end{tabular} ~~~~\\\hline\hline
     \multicolumn{7}{||c||}{\bf Single-Path Programs} \\ \hline\hline
    ~fac & 26 & 0 &  0.35s/6.91MB & 1883 & 4/34 & 26/1.3\%\\ \hline
    ~fib & 74 & 0 &  0.25s/5.68MB & 571 & 4/22 & 26/4.5\%\\ \hline
    ~janne-complex$^\ast$~ & 65 & 0 &  0.54s/7.76MB & 792  & 0/23 &\textbf{176/22\%}\\ \hline
    ~matmult$^\ast$ & 162 & 0 &  119.2s/936.75MB  & 614827 & \textbf{31/107} & 800/0.001\%\\ \hline
    ~jfdcint & 374 & 0 &  7.13s/55.99MB & 49017 & \textbf{394/454} & 108/0.22\% \\ \hline
    ~expint(50,1) & 81  & 0 &  6.08s/59.16MB & 65042 & 0/124 &  70/1.7\% \\ \hline
    ~expint(50,21) & 81  & 0 &  3.65s/43.21MB & 41015 & 0/124 & 71/1.7\%\\ \hline
    ~fdct & 238  & 0 &  2.83s/26.79MB & 26099 & 0/286 & 90/0.3\%\\ \hline
    ~edn$^\ast$ & 284  & 0 &  22.28s/230.98MB & 62968 & 0/460 & 26/0.04\%\\ \hline
    ~recursion$^\ast$ & 41  & 0 &  2.68s/28.82MB & 10335 & 0/38 &  32/0.3\%\\ \hline\hline
    \multicolumn{7}{||c||}{\bf Multiple-Paths Programs} \\ \hline\hline
    ~bs & 174 & 5 &  0.52s/6.52MB  & 366 & 0/22 & \textbf{30/8.2\%} \\ \hline
    ~cnt$^\ast$ & 115 & 100 & 100.25s/377.02MB & 6483 & 0/82 & 40/0.06\% \\ \hline
    ~insertsort$^\ast$ & 91 &  675 &  9.36s/81.27MB & 27061 & 0/53 & 400/1.4\%\\ \hline
    ~ns$^\ast$ & 497 & 625 &  12.38s/110.92MB & 43239 & 0/41 & 32/0.0007\%\\  \hline\hline
  \end{tabular}
  \begin{flushleft}
    \rule{3cm}{.3pt}\\
    $^\dagger$lines of code in the C source file \hfill $^\ddagger N=$ Max number of Player~2 moves along a path \\
    $^\star$Abstracted Instr./Instr. \hfill $^\ast$Program selected for the
    WCET Challenge 2006~\cite{bench-malar}
  \end{flushleft}
%
  \caption{Results (C programs compiled with \texttt{gcc -O2})}
  \label{tab-results}
\end{table}

\vspace{-1.3cm}
\section{Conclusion}

In this paper we have presented a framework based on timed games and
the model checker UPPAAL-TiGA to compute WCET for programs running on
architectures featuring pipelines and caches.

\noindent The results we have obtained support the claim that model
checking is adequate for computing WCET. Moreover UPPAAL-TiGA could be
tuned to handle WCET computation more efficiently: \emph{priorities}
between processes can reduce unnecessary interleavings and there are
not yet implemented in UPPAAL-TiGA (though they are in UPPAAL); a lot
of time is spent \emph{checking} whether a new state has already been
encoutered: this will never be the case in the programs we check
(otherwise they would be an infinite loop). Disabling this check
would also reduce the time to compute the results.  Of course, a
program like Bubble Sort remains beyond the scope of analysis within
our framework. Nevertheless, what we advocate is the combination of
different techniques to solve the WCET problem: \emph{abstract
  interpretation} (AI) combined with \emph{Interger Linear
  Programming} (ILP) have given very good
results~\cite{wcet-ai-aswsd-ferdinand-04} but this method is yet to
prove that: (1) it can be \emph{easily} adapted to different
processors and (2) it can take into account \emph{power} related
features (like change of speed of the processor).

\smallskip

\noindent Our ongoing work focuses on two aspects:
\vspace{-.2cm}
\begin{enumerate}
\item extend the set of instructions supported by our compiler and 
  provide models for other architectures (like ARM11);
\item add a \emph{pre-processing} step to prune the execution tree of
  the program. The goal of this step is to reduce the number of paths
  of the program still preserving the paths giving the WCET.  This
  step can be carried out using ILP techniques, or
  \emph{counter-example guided abstraction refinement} (CEGAR)
  methods~\cite{clarke-cegar-acm-03}.
\end{enumerate}

\paragraph{\bfseries Acknowledgements.} \mbox{}\\
The author would like to thank Bernard Blackham (NICTA, Sydney) and
Gernot Heiser (NICTA, Sydney) for their helpful comments and support.

\bibliography{wcet}

\end{document}